\documentclass{ws-ijbc}
\usepackage{ws-rotating}     
\usepackage{graphicx}
\usepackage{epstopdf}

\begin{document}
\catchline{}{}{}{}{} 

\markboth{M. Katsanikas $\&$ S. Wiggins}{The Generalization of the Periodic Orbit Dividing Surface}
\title{The Generalization of the Periodic Orbit Dividing Surface in Hamiltonian Systems with three  or more  degrees of freedom - I}
\author{MATTHAIOS  KATSANIKAS  AND  STEPHEN  WIGGINS}
\address{School of Mathematics, University of Bristol, Fry Building, Woodland Road, Bristol BS8 1UG, United Kingdom \\ matthaios.katsanikas@bristol.ac.uk, s.wiggins@bristol.ac.uk}

\maketitle

\begin{history}
\received{(to be inserted by publisher)}
\end{history}

\maketitle
\begin{abstract}
We present a method  that generalizes the periodic orbit dividing surface construction for Hamiltonian systems with three or more degrees of freedom. We construct a  torus  using as a basis a periodic orbit and we extend this to a $2n-2$ dimensional object in the $2n-1$ dimensional energy surface. We present  our methods using  benchmark examples for two and  three degree of freedom Hamiltonian systems to illustrate the corresponding algorithm for this construction. 
Towards this end  we use the normal form quadratic Hamiltonian system with two and  three degrees of freedom.  We found that  the periodic orbit dividing surface can provide us the same dynamical information as the dividing surface constructed using normally hyperbolic invariant manifolds. This is significant because, in general, computations of normally hyperbolic invariant manifolds are very difficult in Hamiltonian systems with three or more degrees of freedom. However, our method  avoids this computation and  the only information that we need is the location of one  periodic orbit.
\end{abstract}


\section{Introduction}
\label{intro}
In this paper we construct   dividing surfaces from periodic orbits in Hamiltonian systems with three or more degrees of freedom. This generalises the periodic orbit dividing surface construction developed in \cite{Pechukas73, Pechukas77, Pollak78}.  

A dividing surface (at a fixed value of energy)  is a surface that has one less dimension than that of the energy  surface of a Hamiltonian system. This surface has the property of no-recrossing and orientability. For our purposes the significance of orientability is that a surface has two well-defined sides. We underline the fact that we must not confuse the no-recrossing property of the dividing surfaces  with the Poincar{\'e} recurrence in the case of closed and bounded energy surfaces. In the case of    Poincar{\'e} recurrence the  trajectories  starting from an initial point in the phase space (included the dividing surfaces) will eventually return close to the initial position. This is independent from the choice of the dividing surface. Dividing surfaces play a central role in Wigner's vision for transition state theory \cite{Wigner38, waalkens2007}, which is a widely used theory for understanding chemical reaction dynamics.

In Hamiltonian systems with two degrees of freedom the phase space is four-dimensional and the energy surface is three-dimensional. This means that the dividing surface is two-dimensional surface  embedded in the three-dimensional energy surface. This dividing surface is constructed using periodic orbits (\cite{pechukas1981},  \cite{pollak1985}) that are 1-dimensional closed curves. 

Until now,  the construction of a dividing surface in Hamiltonian systems with three or more degrees  can be done from the Normally Hyperbolic Invariant Manifold (NHIMs)
\cite{wiggins2001, uzer2002, wiggins2016}. The phase space is $2n$-dimensional and the energy surface is  $2n-1$ dimensional surface. The periodic orbit is one-dimensional and it has not enough dimensions to construct a dividing surface using the method that we use in Hamiltonian systems with two degrees of freedom. The Normally Hyperbolic Invariant Manifold is a $2n-3$ dimensional structure in the  $2n-1$ dimensional energy surface. The dimensions of this object can guarantee the construction of a $2n-2$ dimensional structure  embedded in the $2n-1$ dimensional energy surface \cite{waalkens2007}, \cite{waalkens2010}. The construction of a  Dividing surface from the NHIM  can be done through the  computation of the Poincar{\'e}-Birkhoff normal form (NF) theory near an index-one saddle.  By this method we obtain  explicit formulas for the Normally Hyperbolic Invariant Manifolds and Dividing surfaces (\cite{wiggins2001}, \cite{uzer2002}, \cite{waalkens2007}, \cite{toda2003}, \cite{komatsuzaki2003}). But in many situations this method is difficult to apply  and it requires extensive algebraic calculations. 

In this paper we  propose  a new method to construct dividing surfaces without to know the NHIM, avoiding the extensive and difficult calculations through the Normal Form theory. We use the periodic orbit, that is actually a 1-dimensional submanifold of the Normally Hyperbolic Invariant Manifold, as a starting point to  construct a dividing surface. We give an introduction and pseudocode of our algorithm in the general context of the Hamiltonian Systems with $n$ degrees of freedom in the section \ref{sec.1}. We describe our algorithm for the case of the  Hamiltonian systems with two and three degrees of freedom in the sections \ref{sec.2} and \ref{sec.3} respectively. In these sections, we apply our algorithm to a quadratic normal form Hamiltonian system with two and three degrees of freedom and we compare the periodic orbit dividing surfaces that are constructed from our algorithm with the dividing surfaces that are constructed from the NHIM.  We choose this system because we have analytical formulas for the periodic orbits and the NHIM (\cite{ezra2018}). This allows us to compute the dividing surfaces from the periodic orbits and the NHIM and to compare these results (see sections \ref{sec.2} and  \ref{sec.3}). Then we present the algorithm in the general case of a Hamiltonian system with $n$ degrees of freedom (see section \ref{sec.4}). Finally, we present our conclusions in the last section.   

\section{The skeleton of  the Algorithm - Pseudocode}
\label{sec.1}

In this section we  present a  method for the construction of a periodic orbit  dividing surface in Hamiltonian systems with three or more degrees of freedom. This method generalizes  the classical  periodic orbit dividing surface  that is valid only in Hamiltonian systems with two degrees of freedom. We consider  the general case of Hamiltonian systems with $n$ degrees of freedom with potential energy function
$V(x_1,x_2,...,x_n)$ with $n\geq2$ of the form :

\begin{eqnarray}
\label{hgen1}
T +V(x_1,x_2,...,x_n)=E
\nonumber\\
\end{eqnarray}

\noindent
where $T$ is the kinetic energy. 

\begin{eqnarray}
\label{hgen1a}
T=p_{x_1}^2/2m_1+p_{x_2}^2/2m_2+...+p_{x_n}^2/2m_n
\nonumber\\
\end{eqnarray}
where $p_{x_1},p_{x_2},...p_{x_n}$ are the momenta and $m_1,m_2,..m_n$ are the corresponding masses.

A basic assumption of our algorithm is that there exists a two dimensional (2D) subspace in the $2n$ dimensional (D) phase space in which the periodic orbit can be represented by a closed curve. With this assumption the algorithm has two versions.

\begin{description}

\item[First version.] In this version we assume that the periodic orbit is represented by a closed curve in a 2D subspace of the phase space that it is not a subspace of the configuration space (for example in the $(x_1,p_{x_1})$ space). In other words, the 2D subspace in this version has one coordinate corresponding to a configuration space variable and the other corresponding to a phase space variable. We refer to this space as the extended configuration space.

\item[Second version.] In this version we assume that the periodic orbit can be represented as a closed curve in a 2D subspace of the configuration space. For example the 2D subspace is described by the coordinates $(x_1,x_2)$.

\end{description}

\begin{figure}
\centering
\includegraphics[angle=0,width=19.0cm]{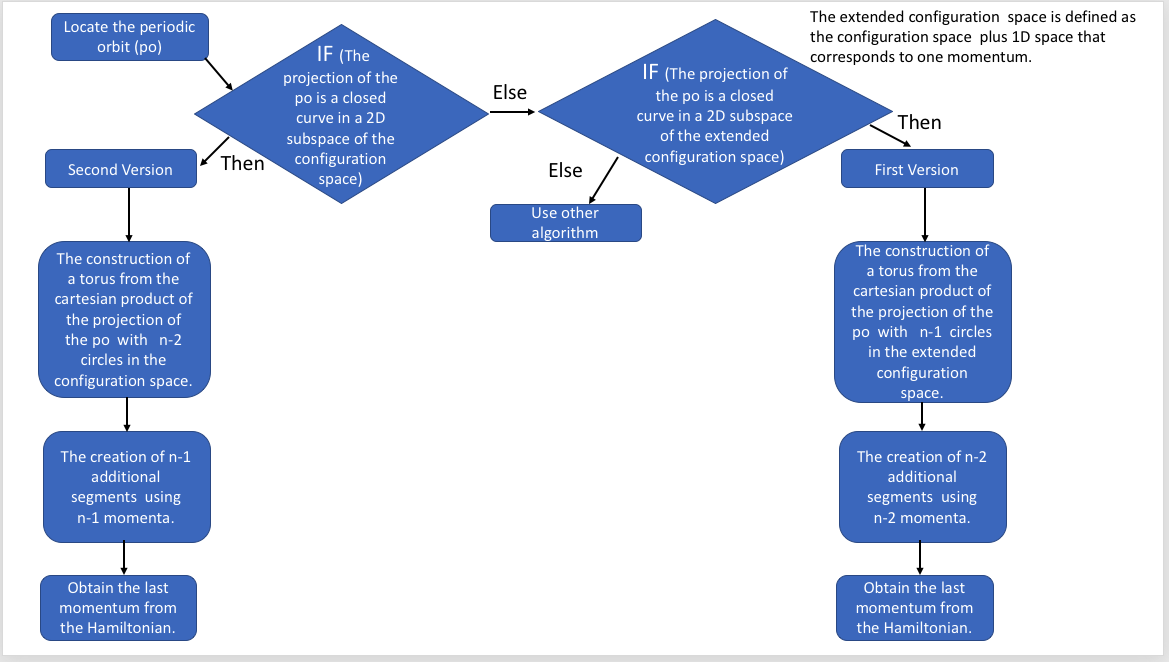}
\caption{The pseudocode for the algorithm.}
\label{code}
\end{figure}

In this section, we describe the two versions of the algorithm for the construction of the dividing surfaces if we know the location of one periodic orbit. For this reason we present a diagram (pseudocode) that describes the two versions of the algorithm (Fig. \ref{code}). We see in this diagram that we check first if the periodic orbit satisfies the conditions that are necessary in order to use one of two versions or not. Firstly, we check if the periodic orbit is valid for the second version. If this is true then we follow the steps of the second version (see Fig. \ref{code}) otherwise we check the validity of the first version of the algorithm. If the periodic orbit satisfies the conditions for the first version then we follow the steps of this version (see Fig. \ref{code}) otherwise we must use another method.

\section{The Algorithm for Hamiltonian systems with two degrees of freedom}
\label{sec.2}
In the  previous section  we described  the difference between  the two versions of our algorithm for the construction of dividing surfaces and the circumstances when we use one or the other version. Now, we describe in detail the two versions  of our algorithm (for $n=2$) in the case of Hamiltonian systems with two degrees of freedom with  potential energy function $V(x,y)$: 

\begin{eqnarray}
\label{hgen}
p_x^2/2m_1+p_y^2/2m_2+V(x,y)=E
\nonumber\\
\end{eqnarray}

\noindent
where $p_x,p_y$ are the momenta and $m_1,m_2$ the corresponding masses.\\
 The algorithm  for Hamiltonian systems 
 with two degrees of freedom  comprises the next two algorithms that correspond to the first and second version of our algorithm. The application of the second version in Hamiltonian systems with two degrees of freedom   give us the same algorithm as the classical algorithm for periodic orbit  dividing surfaces (see \cite{Pechukas73, Pechukas77, Pollak78, Pechukas79, pollak1985} and for more details see \cite{ezra2018}). We describe the two versions of the algorithm in the first two subsections and we present an application of the algorithm to the case of a quadratic normal form Hamiltonian with two degrees of freedom in the third subsection.

\subsection{The first version of the algorithm} 
The first version of the algorithm  is:

\begin{enumerate}

\item Locate an unstable periodic orbit PO for a fixed value of Energy $E$.

\item  Project the PO into the configuration space and  consider  a 2D subspace of the phase space in which the projection of the periodic orbit is a closed curve (for example in the $(y,p_y)$ space). 

\item   From the projection of the periodic orbit in the configuration space, we construct a torus that is generated by the Cartesian product  of a  circle with small radius and the projection of the periodic orbit in a 2D subspace of the phase space (for example in the $(y,p_y)$ space).   Actually it is topologically equivalent to the Cartesian product of two circles $S^{1}\times S^{1}$. This is a two-dimensional torus. This can be achieved through the  construction  of one circle around every point of the   periodic orbit in the 2D subspace of the  3D subspace of the phase space. For example  we  compute a circle (with a fixed radius r) in the plane $(x,y)$ around every point of the periodic orbit in the   3D space  $(x,y,p_y)$.  The goal of this step is to include all coordinates of the configuration space in this torus.

\begin{eqnarray}
\label{eqt2a}
y_{1,i,j1}=y_{0,i} + r cos(\theta_{j1})
\nonumber\\
x_{1,i,j1}=x_{0,i} + r sin(\theta_{j1})
\nonumber\\
p_{y,1,i,j1}=p_{y,0,i}
\end{eqnarray}

\noindent
 $(x_{0,i},y_{0,i},p_{y,0,i}), i=1,...N$ are  the points of the periodic orbit in the 3D subspace $(x,y,p_y)$. We have the angle $\theta_{j1}=j1\frac{2\pi}{n_1}$ with $j1=1,...,n_1$ for the circle  that we need for the construction of the torus. $x_{1,i,j1},y_{1,i,j1},p_{y,1,i,j1}$ with $i=1,...,N$ and $j1=1,...,n_1$  are the points of the torus that is constructed from the Cartesian product of  projection of the periodic orbit in the 2D subspace $(y,p_y)$  and a circle in the $(x,y)$ space in the  3D space  $(x,y,p_y)$.

\item For each point $x_{1,i,j1},y_{1,i,j1},p_{y,1,i,j1}$ on this torus we can  calculate the $p_{x,1,i,j1}$  by  solving the following equation for  a fixed value of energy (Hamiltonian)  $E$:

\begin{eqnarray}
\label{eq6}
V(x_{1,i,j1},y_{1,i,j1})+\frac{p_{x,1,i,j1}^2}{2m_1}+
\frac{p_{y,1,i,j1}^2}{2m_2}=E
\nonumber\\
\end{eqnarray}

\end{enumerate}

\noindent
\textit{Dimensionality and Topology}: This algorithm constructs a torus as the product of the closed curve that represents the projection of the periodic orbit (1D object) in a 2D subspace of the phase space with one circle in the 3D energy surface.  This torus is a 2-dimensional.  Then we obtain the value of the last momentum from the Hamiltonian of the system. 

\subsection{The second version of the algorithm}

The second version of the algorithm is:

\begin{enumerate}

\item Locate an unstable periodic orbit PO for a fixed value of Energy $E$.

\item $(x_{0,i},y_{0,i}), i=1,...N$ are  the points of the periodic orbit in the 2D configuration space $(x,y)$. This is a closed curve in a 2D subspace. Actually it is topologically equivalent with a circle.

\item For each point $x_{0,i},y_{0,i}$ of the periodic orbit we must  calculate the $p_{x,0,i}^{max}$ and   $p_{x,0,i}^{min}$  by  solving the following equation for  a fixed value of energy (Hamiltonian)  $E$ with $p_y=0$:

\begin{eqnarray}
\label{eqt6a}
V(x_{0,i},y_{0,i})+\frac{p_{x,0,i}^2}{2m_1}=E
\nonumber\\
\end{eqnarray}

and we find the maximum and minimum values 
$p_{x,0,i}^{max}$ and   $p_{x,0,i}^{min}$. 
We choose  points $p_{x,0,i}$ in the interval  $p_{x,0,i}^{min}\leq p_{x,0,i}\leq  p_{x,0,i}^{max}$. These points can be uniformly distributed in this interval. 
\item Now for every point $x_{0,i},y_{0,i},p_{x,0,i}$ we must  calculate the $p_{y,0,i}$  by  solving the following equation for  a fixed value of energy (Hamiltonian)  E:

\begin{eqnarray}
\label{eqt6a1}
V(x_{0,i},y_{0,i})+\frac{p_{x,0,i}^2}{2m_1}+
\frac{p_{y,0,i}^2}{2m_2}=E
\nonumber\\
\end{eqnarray}

\end{enumerate}

\noindent
\textit{Dimensionality and Topology}: This algorithm give us an 1-dimensional object (a circle or ellipse)  as the projection of the periodic orbit (1D object) in  the  configuration space.  Then we sample the third variable (one of the momenta)  in the interval between its  maximum and minimum value. Actually we create an additional 1D segment and we increase  the dimensionality of the initial object (a circle or ellipse), from 1 to 2 dimensions, which is  embedded in the 3D  energy surface. Then we obtain the value of the last momentum from the Hamiltonian of the system.

\subsection{Application of the algorithm to the Quadratic Normal Form Hamiltonian System with two degrees of freedom} 
\label{sec.2a}

Now we will apply the  algorithm to the system that is described in subsection \ref{modela}. In this system the reaction occurs when $x$ changes sign \cite{ezra2018}. The dividing surfaces for this system  using the classical algorithm have been computed (see for example \cite{ezra2018}). In this subsection we will use our algorithm  to  compute the dividing surfaces.
This can be done  if we choose the condition $x=0$  that is actually a 3-dimensional surface in the four dimensional phase space. This will be applied in our algorithm.  We will construct a dividing surface from the periodic orbit PO that is a circle in the plane $(y,p_y)$ (see subsection \ref{sub1a}). The first version of our algorithm is valid for this periodic orbit because we have the combination of one coordinate of the configuration space and one momentum (see for details the diagram in section \ref{sec.1}). We underline the fact that because one of the coordinates of the configuration space is $x=0$ (in our case) we have not the need to construct a two-dimensional torus. This means that the first version of the  algorithm  is simplified and we construct the torus using the one circle that is the projection of the periodic orbit. 

\subsubsection{Hamiltonian Model}
\label{modela}

The quadratic normal form Hamiltonian system near an index-one saddle \cite{wiggins2016} is described by the following Hamiltonian: 

\begin{eqnarray}
\label{eq1a}
H=\frac{\lambda}{2} (p_x^2 - x^2) + \frac{\omega}{2}(p_y^2+y^2) 
\nonumber\\
\end{eqnarray}

with $\lambda >0,\omega>0$ and 

\begin{eqnarray}
\label{eq2a}
H_1=\frac{\lambda}{2} (p_x^2 - x^2) 
\nonumber\\
H_2=\frac{\omega}{2}(p_y^2+y^2)  
\nonumber\\
\end{eqnarray}

The equations of motion  are:
 
\begin{eqnarray}
\label{eq3a}
 \dot x= \frac{\partial H}{\partial p_x}=\lambda p_x
 \nonumber\\
 \dot p_x =-\frac{\partial H}{\partial x}=\lambda x
 \nonumber\\
 \dot y =\frac{\partial H}{\partial p_y}=\omega p_y
 \nonumber\\
 \dot p_y =-\frac {\partial H} {\partial y}=-\omega y
 \nonumber\\ 
\end{eqnarray}

The Hamiltonian functions $H_1$ and  $H_2$ are uncoupled. The phase space of these two subsystems   can be studied separately.  In this system we can distribute the total energy between each mode. The ($x,p_x,y,p_y)=(0,0,0,0)$)  for E=0 is an one-index  saddle point of the full system. 

In this system  the reaction occurs when the $x$ coordinate changes sign (for $H_1>0$). We choose the condition $x=0$ (\cite{ezra2018}) to define a three-dimensional  surface in the four-dimensional phase space. From the equations \ref{eq2a} we have that  $H_2\geq 0$ and  finally  $H=H_1+H_2\equiv E>0$. The energy surface is:

\begin{eqnarray}
\label{eq41}
\frac{\lambda}{2} (p_x^2 - x^2) + \frac{\omega}{2}(p_y^2+y^2) =E
\nonumber\\
\end{eqnarray}

\noindent
the intersection of $x=0$ with this surface is: 

\begin{eqnarray}
\label{eq4a1}
\frac{\lambda}{2} p_x^2  + \frac{\omega}{2}(p_y^2+y^2) =E
\nonumber\\
\end{eqnarray}

This is the Dividing surface (a two-sphere)  that can be constructed by the NHIM (see for more details in \cite{wiggins2016} and \cite{ezra2018}). The NHIM is the unstable periodic orbit in Hamiltonian systems with two degrees of freedom. We call this periodic orbit as PO. The equator of this dividing surface is given by $p_x=0$ and we have: 

\begin{eqnarray}
\label{eq4b1}
 \frac{\omega}{2}(p_y^2+y^2) =E   \qquad PO
\nonumber\\
\end{eqnarray}

This equation represents the  periodic orbit (PO) that is a circle in the plane $(y,p_y)$.In the next section we  will  construct, using the algorithm 1, a two dimensional dividing surface using this periodic orbit.

\subsubsection{PO-Dividing Surface}
\label{sub1a}

We apply the first version (see the introduction of this section) of our algorithm to the PO (see \ref{eq4b1}):

\begin{enumerate}
\item The periodic orbit 1 is given by \ref{eq4b1} for every fixed value $E$.

\item The periodic orbit is a circle and it lies on a plane ($(y,p_y)$. The radius of this circle is  $\sqrt{\frac{2E}{\omega}}$. 

\item  As we described in the introduction of this section  the dividing surface is  embedded in the surface $x=0$ and this means that  we have not the need to construct a two-dimensional torus. This means that this step will be simplified and we will construct 1-dimensional  torus using one circle that is the projection of the periodic orbit in the $(y,p_y)$ plane. The equation of this torus is the equation \ref{eq4b1}.



\item Step 4 is carried out naturally  in this case (with $x=0$). We  compute  
$p_{x}$  from the Hamiltonian.

\noindent
\textit{no-recrossing property}:

\begin{eqnarray}
\label{eq8h}
p_{x}=\sqrt{\frac{2}{\lambda}(H_1+H_2 -\frac{\omega}{2} (p_{y}^2+y^2))} \qquad p_x>0  \qquad Forward \quad  DS 
\nonumber\\
p_{x}= -\sqrt{\frac{2}{\lambda}(H_1+H_2 -\frac{\omega}{2} (p_{y}^2+y^2))} \qquad p_x<0  \qquad Backward \quad DS
\nonumber\\
\end{eqnarray}
\end{enumerate}

The new  DS that is  constructed has the no-recrossing  property because $\dot x=\lambda p_x$. This is exactly the same dividing surface that we obtain from the classical algorithm of \cite{pechukas1981},\citep{pollak1985} (see equations 12 in \cite{ezra2018}).

\section{The Algorithm for Hamiltonian systems with three degrees of freedom}
\label{sec.3}

In the  section \ref{sec.1}  we described  the difference between  the two versions of our algorithm for the construction of dividing surfaces and when we use both versions. Now, we describe in detail the two versions  of our algorithm (for $n=3$) in the case of Hamiltonian systems with three degrees of freedom with a potential energy  $V(x,y,z)$:

\begin{eqnarray}
\label{hgen}
p_x^2/2m_1+p_y^2/2m_2+p_z^2/2m_3+V(x,y,z)=E
\nonumber\\
\end{eqnarray}

\noindent
where $p_x,p_y,p_z$ are the momenta and $m_1,m_2,m_3$ the corresponding masses.\\
When we apply the two versions  of the algorithm  in Hamiltonian systems with three degrees of freedom we produce two algorithms, one algorithm for each version.
We describe the two versions of the algorithm in the first two subsections and we present an application of the algorithm to the case of a quadratic normal form Hamiltonian with three degrees of freedom in the third subsection.

\subsection{The first version of the algorithm} 

The first version of the  algorithm is:

\begin{enumerate}

\item Locate an unstable periodic orbit PO for a fixed value of Energy $E$.

\item  Project the PO into the configuration space and we consider  a 2D subspace of the phase space in which the projection of the periodic orbit is a closed curve (for example in the $(y,p_y)$ space). 

\item   From the projection of the periodic orbit in the configuration space, we construct a torus that is generated by the Cartesian product  of 2  circles with small radius and the projection of the periodic orbit in a 2D subspace of the phase space (for example in the $(y,p_y)$ space).   Actually it is topologically equivalent with the Cartesian product of three circles $S^{1}\times S^{1}\times S^{1}$. This is a three-dimensional torus (Hypertorus). This can be achieved through the  construction  of one circle around every point of the   periodic orbit in the 2D subspace of a 4D subspace of the phase space. For example  we  compute a circle (with a fixed radius r) in the plane $(x,y)$ around every point of the periodic orbit in the   4D subspace  $(x,y,z,p_y)$. Then we construct a new circle  around every point of the previous structure in the 4D subspace $(x,y,z,p_y)$ of the phase space. This can be done computing a circle (with a  fixed radius r) in the plane $(y,z)$. The goal of this step is to include all coordinates of the configuration space in this torus.

\begin{eqnarray}
\label{eqt2}
y_{1,i,j1}=y_{0,i} + r cos(\theta_{j1})
\nonumber\\
x_{1,i,j1}=x_{0,i} + r sin(\theta_{j1})
\nonumber\\
z_{1,i,j1}=z_{0,i}
\nonumber\\
p_{y,1,i,j1}=p_{y,0,i}
\end{eqnarray}

\begin{eqnarray}
\label{eqt2a}
y_{2,i,j1,j2}=y_{1,i,j1}+r cos(\theta_{j2})
\nonumber\\
z_{2,i,j1,j2}=z_{1,i,j1}+r sin(\theta_{j2})
\nonumber\\
x_{2,i,j1,j2}=x_{1,i,j1}
\nonumber\\
p_{y,2,i,j1,j2}=p_{y,1,i,j1}
\end{eqnarray}

 $(x_{0,i},y_{0,i},z_{0,i},p_{y,0,i}), i=1,...N$ are  the points of the periodic orbit in the 4D subspace $(x,y,z,p_y)$. We have the angle $\theta_{j1}=j1\frac{2\pi}{n_1}$ with $j1=1,...,n_1$ for the first circle and  $\theta_{j2}=j2\frac{2k\pi}{n_1}$ with $j2=1,...,n_1$ for the second circle that we need for the construction of the torus.

$x_{1,i,j1},y_{1,i,j1},z_{1,i,j1},p_{y,1,i,j1}$ with $i=1,...,N$ and $j1=1,...,n_1$  are the points of the torus that is constructed from the Cartesian product of  projection of the periodic orbit in the 2D subspace $(y,p_y)$  and a circle in the $(x,y)$ space in the  4D space  $(x,y,z,p_y)$. $x_{2,i,j1,j2},y_{2,i,j1,j2},z_{2,i,j1,j2},p_{y,2,i,j1,j2}$ with  $i=1,...,N$, $j1=1,...,n_1$  and $j2=1,...,n_1$ are the points of the torus that is constructed from the Cartesian product of the projection of the periodic orbit in the 2D subspace $(y,p_y)$, a circle in the $(x,y)$ space and a circle in the $(y,z)$ space in the  4D space $(x,y,z,p_y)$.

\item For each point $x_{2,i,j1,j2},y_{2,i,j1,j2},z_{2,i,j1,j2},p_{y,2,i,j1,j2}$ on this torus we must  calculate the $p_{x,2,i,j1,j2}^{max}$ and   $p_{x,2,i,j1,j2}^{min}$  by  solving the following equation for  a fixed value of energy (Hamiltonian)  $E$ with $p_z=0$:

\begin{eqnarray}
\label{eq6}
V(x_{2,i,j1,j2},y_{2,i,j1,j2},z_{2,i,j1,j2})+\frac{p_{x,2,i,j1,j2}^2}{2m_1}+
\frac{p_{y,2,i,j1,j2}^2}{2m_2}=E
\nonumber\\
\end{eqnarray}

\noindent
and we find the maximum and minimum values   
$p_{x,2,i,j1,j2}^{max}$ and   $p_{x,2,i,j1,j2}^{min}$. We choose  points $p_{x,2,i,j1,j2}$ with $j2=1,...,n_1$ in the interval  $p_{x,2,i,j1,j2}^{min}\leq p_{x,2,i,j1,j2}\leq  p_{x,2,i,j1,j2}^{max}$. These points can be uniformly distributed in this interval. Then we obtain the value  $p_{z,2,i,j1,j2}$ from the Hamiltonian:

\begin{eqnarray}
\label{eqt6}
V(x_{2,i,j1,j2},y_{2,i,j1,j2},z_{2,i,j1,j2})+\frac{p_{x,2,i,j1,j2}^2}{2m_1}+
\frac{p_{y,2,i,j1,j2}^2}{2m_2}+\frac{p_{z,2,i,j1,j2}^2}{2m_3}=E
\nonumber\\
\end{eqnarray}

\end{enumerate}

\noindent
\textit{Dimensionality and Topology}: This algorithm constructs a torus as the product of the closed curve that represents the projection of the periodic orbit (1D object) in a 2D subspace with two circles in the 4D subspace of the 5D energy manifold.  This torus is a 3-dimensional (hypertorus). Then we sample the fifth variable (one of the momenta)  in the interval between its  maximum and minimum value. Actually we create an additional 1D segment and we increase  the dimensionality of the initial torus, from 3 to 4 dimensions, which is  embedded in the 5D energy surface. Then we obtain the value of the last momentum from the Hamiltonian of the system. 

\subsection{The second version of the algorithm}

The second version of the algorithm is:

\begin{enumerate}

\item Locate an unstable periodic orbit PO for a fixed value of Energy $E$.

\item Project the PO into the configuration space and we consider   a 2D subspace of the configuration space  in which the projection of the periodic orbit is a closed curve (for example in the $(x,y)$ space). 

\item  We construct a torus that is generated by the Cartesian product  of one circle with small radius and the projection of the periodic orbit in a 2D subspace of the configuration space (for example in the $(x,y)$ space). Actually it is the Cartesian product of two circles $ S^{1}\times S^{1}$. This is a two-dimensional torus. This can be achieved through the  construction  of one circle around every point of the   periodic orbit in the 2D subspace of  a 3D subspace of the phase space. For example  we  compute a circle (with a fixed radius r) in the plane $(y,z)$ around every point of the periodic orbit in the 3D subspace  $(x,y,z)$. 

The points of the torus that we constructed  are:

\begin{eqnarray}
\label{eqt3}
y_{1,i,j1}=y_{0,i}+r cos(\theta_{j1})
\nonumber\\
z_{1,i,j1}=z_{0,i}+r sin(\theta_{j1})
\nonumber\\
x_{1,i,j1}=x_{0,i}
\end{eqnarray}

 $(x_{0,i},y_{0,i},z_{0,i}), i=1,...N$ are  the points of the periodic orbit in the 3D configuration space $(x,y,z)$. We have the angle $\theta_{j1}=j1\frac{2\pi}{n_1}$ with $j1=1,...,n_1$ for the  circle  that we need for the construction of the torus.

$x_{1,i,j1},y_{1,i,j1},z_{1,i,j1}$ with $i=1,...,N$ and $j1=1,...,n_1$  are the points of the torus that is constructed from the Cartesian product of  projection of the periodic orbit in the 2D subspace $(x,y)$  and a circle in the $(y,z)$ space in the  3D space $(x,y,z)$. 

\item For each point $x_{1,i,j1},y_{1,i,j1},z_{1,i,j1}$ on this torus we must  calculate the $p_{x,1,i,j1}^{max}$ and   $p_{x,1,i,j1}^{min}$  by  solving the following equation for  a fixed value of energy (Hamiltonian)  $E$ with $p_y=p_z=0$:

\begin{eqnarray}
\label{eqt6a}
V(x_{1,i,j1},y_{1,i,j1},z_{1,i,j1})+\frac{p_{x,1,i,j1}^2}{2m_1}=E
\nonumber\\
\end{eqnarray}

\noindent
and we find the maximum and minimum values 
$p_{x,1,i,j1}^{max}$ and   $p_{x,1,i,j1}^{min}$. 
We choose  points $p_{x,1,i,j1}$ with $j1=1,...,n_1$ in the interval  $p_{x,1,i,j1}^{min}\leq p_{x,1,i,j1}\leq  p_{x,1,i,j1}^{max}$. These points can be uniformly distributed in this interval. 
\item Now for every point $x_{1,i,j1},y_{1,i,j1},z_{1,i,j1},p_{x,1,i,j1}$ we must  calculate the $p_{y,1,i,j1}^{max}$ and   $p_{y,1,i,j1}^{min}$  by  solving the following equation for  a fixed value of energy (Hamiltonian)  $E$ with $p_z=0$:

\begin{eqnarray}
\label{eqt6a1}
V(x_{1,i,j1},y_{1,i,j1},z_{1,i,j1})+\frac{p_{x,1,i,j1}^2}{2m_1}+
\frac{p_{y,1,i,j1}^2}{2m_2}=E
\nonumber\\
\end{eqnarray}

We choose  points $p_{y,1,i,j1}$ with $j1=1,...,n_1$ in the interval  $p_{y,1,i,j1}^{min}\leq p_{y,1,i,j1}\leq  p_{y,1,i,j1}^{max}$. These points can be uniformly distributed in this interval. 
Then we obtain the value  $p_{z,1,i,j1}$ from the Hamiltonian:

\begin{eqnarray}
\label{eqt6a2}
V(x_{1,i,j1},y_{1,i,j1},z_{1,i,j1})+\frac{p_{x,1,i,j1}^2}{2m_1}+
\frac{p_{y,1,i,j1}^2}{2m_2}+\frac{p_{z,1,i,j1}^2}{2m_3}=E
\nonumber\\
\end{eqnarray}

\end{enumerate}

\noindent
\textit{Dimensionality and Topology}: This algorithm constructs a torus as the product of the closed curve that represents the projection of the periodic orbit (1D object) in a 2D subspace with one circle   in the  configuration space.  This torus is a 2-dimensional. Then we sample the fourth variable (one of the momenta)  in the interval between its  maximum and minimum value. Actually we create an additional 1D segment and we increase  the dimensionality of the initial torus, from 2 to 3 dimensions, which is  embedded in the 4D subspace of the 5D energy surface. Then we sample the fifth variable (one of the other two momenta)  in the interval between its  maximum and minimum value. This means that we create an additional 1D segment and we increase  the dimensionality of the initial torus, from 3 to 4 dimensions, which is  embedded in the  5D energy surface. Then we obtain the value of the last momentum from the Hamiltonian of the system.

\subsection{Application of the algorithms  in the Quadratic Normal Form Hamiltonian System with three degrees of freedom} 
\label{sec.3a}

Now we will apply the algorithm to the system that is described in subsection \ref{model}. In this system the reaction occurs when $x$ changes sign \cite{ezra2018}. This can be done  if we choose the condition $x=0$  that is actually a 5-dimensional surface in the six dimensional phase space. This will be applied in our algorithms (in this way, we consider the intersection of the four dimensional structure that is obtained by our algorithm with $x=0$). Firstly  we will construct a dividing surface from the periodic orbit PO1 that is a circle in the plane $(y,p_y)$ (see subsection \ref{sub1}). Then we will do the same for the periodic orbit PO2 that is a circle in the plane $(z,p_z)$
(see subsection \ref{sub2}). In this case we will use  version 1 of the algorithm  because the periodic orbits are closed curves in the 2D subspaces of the phase space that consist of a combination of a coordinate of the configuration space with one momentum (see the conditions for the choice between the first and second version in the section \ref{sec.1}) . We underline the fact that because one of the coordinates of the configuration space is $x=0$ (in our case) we have not the need to construct a three-dimensional torus. This means that the first  version  of the  algorithm is simplified and we construct the torus using the Cartesian product of one circle (not two circles) with the projection of the periodic orbit.

\subsubsection{Hamiltonian Model}
\label{model}

The quadratic normal form Hamiltonian system near an index-one saddle \cite{wiggins2016} is described by the following Hamiltonian: 

\begin{eqnarray}
\label{eq1}
H=\frac{\lambda}{2} (p_x^2 - x^2) + \frac{\omega_2}{2}(p_y^2+y^2) + \frac{\omega_3}{2} (p_z^2+z^2)
\nonumber\\
\end{eqnarray}

\noindent
with $\lambda >0,\omega_2>0,\omega_3>0$ and 

\begin{eqnarray}
\label{eq2}
H_1=\frac{\lambda}{2} (p_x^2 - x^2) 
\nonumber\\
H_2=\frac{\omega_2}{2}(p_y^2+y^2)  
\nonumber\\
H_3=\frac{\omega_3}{2} (p_z^2+z^2)
\nonumber\\
\end{eqnarray}

\noindent
The equations of motion  are:
 
\begin{eqnarray}
\label{eq3}
 \dot x= \frac{\partial H}{\partial p_x}=\lambda p_x 
 \nonumber\\
 \dot p_x =-\frac{\partial H}{\partial x}=\lambda x
 \nonumber\\
 \dot y =\frac{\partial H}{\partial p_y}=\omega_2 p_y
 \nonumber\\
 \dot p_y =-\frac {\partial H} {\partial y}=-\omega_2 y
 \nonumber\\ 
 \dot z =\frac {\partial H} {\partial p_z}=\omega_3 p_z
 \nonumber\\
 \dot p_z =-\frac {\partial H} {\partial z}=-\omega_3 z
 \nonumber\\ 
\end{eqnarray}

\noindent
The Hamiltonian functions $H_1$ ,$H_2$ and $H_3$ are uncoupled. The phase space of these three subsystems   can be studied separately.  In this system we can distribute the total energy between each mode. The ($x,p_x,y,p_y,z,p_z)=(0,0,0,0,0,0)$)  for E=0 is an one-index  saddle point of the full system. 

In this system  the reaction occurs when the $x$ coordinate changes sign (for $H_1>0$). We choose the condition $x=0$ (\cite{ezra2018}) to define a five-dimensional  surface in the six-dimensional phase space. From the equations \ref{eq2} we have that  $H_2\geq 0$ and $H_3\geq 0$ and  finally  $H=H_1+H_2+H_3\equiv E>0$. The energy surface is:

\begin{eqnarray}
\label{eq4}
\frac{\lambda}{2} (p_x^2 - x^2) + \frac{\omega_2}{2}(p_y^2+y^2) + \frac{\omega_3}{2} (p_z^2+z^2)=E
\nonumber\\
\end{eqnarray}

\noindent
the intersection of $x=0$ with this surface is: 

\begin{eqnarray}
\label{eq4a}
\frac{\lambda}{2} p_x^2  + \frac{\omega_2}{2}(p_y^2+y^2) + \frac{\omega_3}{2} (p_z^2+z^2)=E
\nonumber\\
\end{eqnarray}

\noindent
This is the dividing surface (a four-sphere)  that can be constructed by the NHIM (see for more details in \cite{wiggins2016} and \cite{ezra2018}). The equator of this dividing  is given by $p_x=0$  and we have: 

\begin{eqnarray}
\label{eq4b}
 \frac{\omega_2}{2}(p_y^2+y^2) + \frac{\omega_3}{2} (p_z^2+z^2)=E   \qquad NHIM
\nonumber\\
\end{eqnarray}

\noindent
The previous equation represents the NHIM.
The intersection of the NHIM with $z=0$ is given by the following equation:

\begin{eqnarray}
\label{eq4c}
 \frac{\omega_3}{2} p_z^2+ \frac{\omega_2}{2} (p_y^2+y^2)=E  
\nonumber\\
\end{eqnarray}

\noindent
The equator of the NHIM is given by $p_z=0$ and we have the following equation: 

\begin{eqnarray}
\label{eq4d}
 \frac{\omega_2}{2} (p_y^2+y^2)=E   \qquad PO1
\nonumber\\
\end{eqnarray}

\noindent
This equation represents the  periodic orbit (PO1) that is a circle in the plane $(y,p_y)$. 

The intersection of the NHIM (see the equation \ref{eq4b}) with $y=0$  is given by the following equation: 

\begin{eqnarray}
\label{eq4e}
 \frac{\omega_2}{2} p_y^2+ \frac{\omega_3}{2} (p_z^2+z^2)=E  
\nonumber\\
\end{eqnarray}

\noindent
The equator of the NHIM is given by $p_y=0$ and we have the following equation: 

\begin{eqnarray}
\label{eq4f}
 \frac{\omega_3}{2} (p_z^2+z^2)=E  \qquad PO2
\nonumber\\
\end{eqnarray}

This equation represents the  periodic orbit (PO2) that is a circle in the plane $(z,p_z)$. In the next subsections we  will  construct  dividing surfaces using the periodic orbits PO1 and PO2.

\subsubsection{PO1-Dividing Surface}
\label{sub1}

We begin by applying  the first  version  of our algorithm  to the PO1 (see \ref{model}):

\begin{enumerate}

\item The PO1 is given by \ref{eq4d} for every fixed value $E$.

\item The periodic orbit is a circle and it lies on a plane ($(y,p_y)$. The radius of this circle is  $\sqrt{\frac{2E}{\omega_2}}$. 

\item  We construct a torus in the 3D subspace $(y,p_y,z)$  using the product of the circle of the periodic orbit and a circle with small fixed radius $r>0$. The equation of this torus (with $x=0$):

\begin{eqnarray}
\label{eq8a}
(\sqrt{p_{y}^2+y^2} - \sqrt{\frac{2E}{\omega_2}})^2 + z^2=r^2
\nonumber\\
\end{eqnarray}



\item Step 4 is carried out naturally  in this case (with $x=0$). We  compute the 
$p_{z}^{max}$ and $p_{z}^{min}$. We sample points in the interval 
$[p_{z}^{min}, p_{z}^{max}]$. For every point in the previous interval we  obtain  the $p_{x}$ coordinate  from the Hamiltonian.

\noindent
\textit{no-recrossing property}: 

from the equation \ref{eq8a} we have: 

\begin{eqnarray}
\label{eq8c}
z^2=r^2-(\sqrt{p_{y}^2+y^2} - \sqrt{\frac{2E}{\omega_2}})^2
\nonumber\\
\nonumber\\
\end{eqnarray}


\noindent
from the equation \ref{eq1}  (for a fixed value of energy $E$, the numerical value of the Hamiltonian and for $x=0$)  we have:

\begin{eqnarray}
\label{eq8e}
\frac{\lambda}{2} p_{x}^2=E-\frac{\omega_2}{2} (p_{y}^2+y^2)-\frac{\omega_3}{2} (p_{z}^2+z^2)
\nonumber\\
\end{eqnarray}

\noindent
The equation \ref{eq8e} through the equation \ref{eq8c}
and  using that $E=H_1+H_2+H_3$ with $H_2=\frac{\omega_2}{2} (p_{y}^2+y^2)$ (from equation \ref{eq2}) we have:

\begin{eqnarray}
\label{eq8f}
\frac{\lambda}{2} p_{x}^2=H_1+H_3 -\frac{\omega_3}{2}p_{z}^2+\frac{\omega_3}{2} (\sqrt{p_{y}^2+y^2} - \sqrt{\frac{2E}{\omega_2}})^2-\frac{\omega_3}{2} r^2
\nonumber\\
\end{eqnarray}

\noindent
From equation \ref{eq2} we have that $H_3-\frac{\omega_3}{2}p_{z}^2>0$. We have also that the quantity $\frac{\omega_3}{2} (\sqrt{p_{y}^2+y^2} - \sqrt{\frac{2E}{\omega_2}})^2 \geq 0$ is always positive. This means that the condition that we must have for the equation \ref{eq8f} is:

\begin{eqnarray}
\label{eq8g}
H_1-\frac{\omega_3}{2} r^2 \geq 0
\nonumber\\
\end{eqnarray}

\noindent
This implies that:

\begin{eqnarray}
\label{eq8g2}
r \leq \sqrt{\frac{2H_1}{\omega_3}}
\nonumber\\
\end{eqnarray}

In practise this means that we must choose a small radius r.  Using this condition we have for the dividing surface :

\begin{eqnarray}
\label{eq8h}
p_{x}=\sqrt{\frac{2}{\lambda}(H_1+H_3 -\frac{\omega_3}{2}p_{z}^2+\frac{\omega_3}{2} (\sqrt{p_{y}^2+y^2} - \sqrt{\frac{2E}{\omega_2}})^2-\frac{\omega_3}{2} r^2)} \qquad p_x>0  \qquad Forward \quad  DS 
\nonumber\\
p_{x}=-\sqrt{\frac{2}{\lambda}(H_1+H_3-\frac{\omega_3}{2}p_{z}^2+\frac{\omega_3}{2} (\sqrt{p_{y}^2+y^2} - \sqrt{\frac{2E}{\omega_2}})^2-\frac{\omega_3}{2} r^2)} \qquad p_x<0  \qquad Backward \quad DS
\nonumber\\
\end{eqnarray}
\end{enumerate}

\noindent
The new  DS that is  constructed has the no-recrossing  property because $\dot x=\lambda p_x$.

\subsubsection{PO2-Dividing Surface}
\label{sub2}

Now we apply the first version of our algorithm to the PO2 (see \ref{model}):

\begin{enumerate}

\item The PO2 is given by \ref{eq4f} for every fixed value $E$.

\item The periodic orbit is a circle and it lies on a plane $(z,p_z)$. The radius of this circle is  $\sqrt{\frac{2E}{\omega_3}}$. 

\item  We construct a torus in the 3D subspace $(z,p_z,y)$  using the product of the circle of the periodic orbit and a circle with small fixed radius $r_1>0$. The equation of this torus (with $x=0$):

\begin{eqnarray}
\label{eq8a2}
(\sqrt{p_{z}^2+z^2} - \sqrt{\frac{2E}{\omega_3}})^2 + y^2=r_1^2
\nonumber\\
\end{eqnarray}



\item Step 4 is carried out naturally  in this case (with $x=0$). We  compute the 
$p_{y}^{max}$ and $p_{y}^{min}$. We sample points in the interval 
$[p_{y}^{min}, p_{y}^{max}]$. For every point in the previous interval we  obtain  the $p_{x}$ coordinate  from the Hamiltonian. 

\noindent
\textit{no-recrossing property}: 

from the equation \ref{eq8a2} we have: 

\begin{eqnarray}
\label{eq8c2}
y^2=r_{1}^2-(\sqrt{p_{z}^2+z^2} - \sqrt{\frac{2E}{\omega_3}})^2
\nonumber\\
\nonumber\\
\end{eqnarray}



\noindent
from the equation \ref{eq1}  (for a fixed value of energy $E$, the numerical value of the Hamiltonian)  we have:

\begin{eqnarray}
\label{eq8e2}
\frac{\lambda}{2} p_{x}^2=E-\frac{\omega_2}{2} (p_{y}^2+y^2)-\frac{\omega_3}{2} (p_{z}^2+z^2)
\nonumber\\
\end{eqnarray}

\noindent
The equation \ref{eq8e2} through the equation \ref{eq8c2}
and  using that $E=H_1+H_2+H_3$ with $H_3=\frac{\omega_3}{2} (p_{z}^2+z^2)$ (from equation \ref{eq2}) we have:

\begin{eqnarray}
\label{eq8f2}
\frac{\lambda}{2} p_{x}^2=H_1+H_2 -\frac{\omega_2}{2}p_{y}^2+\frac{\omega_2}{2} (\sqrt{p_{z}^2+z^2} - \sqrt{\frac{2E}{\omega_3}})^2-\frac{\omega_2}{2} r_{1}^2
\nonumber\\
\end{eqnarray}

\noindent
From equation \ref{eq2} we have that $H_2-\frac{\omega_2}{2}p_{y}^2>0$. We have also that the quantity $\frac{\omega_2}{2} (\sqrt{p_{y}^2+y^2} - \sqrt{\frac{2E}{\omega_3}})^2 \geq 0$ is always positive. This means that the condition that we must have for the equation \ref{eq8f2} is:

\begin{eqnarray}
\label{eq8g2b}
H_1-\frac{\omega_2}{2} r_{1}^2 \geq 0
\nonumber\\
\end{eqnarray}

This implies that:

\begin{eqnarray}
\label{eq8g22}
r_{1} \leq \sqrt{\frac{2H_1}{\omega_2}}
\nonumber\\
\end{eqnarray}

\noindent
In practise this means that we must choose a small radius $r_1$.  Using this condition we 
have for the dividing surface :

\begin{eqnarray}
\label{eq8h2}
p_{x}=\sqrt{\frac{2}{\lambda}(H_1+H_2 -\frac{\omega_2}{2}p_{y}^2+\frac{\omega_2}{2} (\sqrt{p_{z}^2+z^2} - \sqrt{\frac{2E}{\omega_3}})^2-\frac{\omega_2}{2} r_{1}^2)} \qquad p_x>0  \qquad Forward \quad  DS 
\nonumber\\
p_{x}=-\sqrt{\frac{2}{\lambda}(H_1+H_2-\frac{\omega_2}{2}p_{y}^2+\frac{\omega_2}{2} (\sqrt{p_{z}^2+z^2} - \sqrt{\frac{2E}{\omega_3}})^2-\frac{\omega_2}{2} r_{1}^2)} \qquad p_x<0  \qquad Backward \quad DS
\nonumber\\
\end{eqnarray}
\end{enumerate}

\noindent
The new  DS that is  constructed has the no-recrossing  property because $\dot x=\lambda p_x$. 

\subsubsection{The structure and the comparison of the Dividing surfaces }
\label{sec.3s}

In this section we constructed  the dividing surfaces  from the periodic orbits  PO1 and PO2 and from the Normally Hyperbolic Invariant Manifold (NHIM) of the quadratic normal form Hamiltonian (see section \ref{model}) for ($\lambda=1,\omega_2= \sqrt{2},\omega_3=1$). These values were used from \cite{ezra2018}. The value of the energy was  $E=14, H_1=4$.  We computed the dividing surfaces from the periodic orbits PO1 and PO2 using the algorithms and  analytical formulas of the previous subsections (see subsections \ref{sub1} and \ref{sub2}).  The dividing surface from the NHIM is obtained using the algorithm and analytical formulae  from  \cite{ezra2018}. 

We begin by constructing  the dividing surface from the periodic orbit 1 (PO1) in three different cases. For this purpose we construct a  torus (using as a starting point  the periodic orbit PO1- see subsection \ref{sub1}) that is   the cartesian product of the periodic orbit with a circle. The maximum radius of this circle is $Rmax=\sqrt{\frac{2H_1}{\omega_3}}$ (see the equation \ref{eq8g2}). We  used  three cases for the construction  of a torus  using  a circle with small radius r ($r=Rmax/20$), a circle with larger radius r ($r=Rmax/2$) and a circle with the maximum radius r ($r=Rmax$).

Next we construct  the dividing surface from the periodic orbit 2 (PO2).  For this purpose we construct a  torus (using as a starting point  the periodic orbit PO2- see subsection \ref{sub2}) that is   the cartesian product of the periodic orbit with a circle. The maximum radius $r_{1}$ of this circle is $R1max=\sqrt{\frac{2H_1}{\omega_2}}$ (see the equation \ref{eq8g22}). In this paper we used the maximum radius $r_{1}$  for the construction of the PO2 dividing surface. Finally we construct the dividing surface from the NHIM.

All dividing surfaces as we mentioned in the previous sections  are 4 dimensional  structures  in the energy surface. All dividing surfaces have $x=0$ (as we mentioned   in the previous section and  we  obtain the $p_x$ coordinate from the Hamiltonian) and this is the reason that we studied the structure of the dividing surfaces  in the  4-dimensional space $(y,z,p_y,p_z)$.

\begin{figure}
\centering
\includegraphics[angle=0,width=5.0cm]{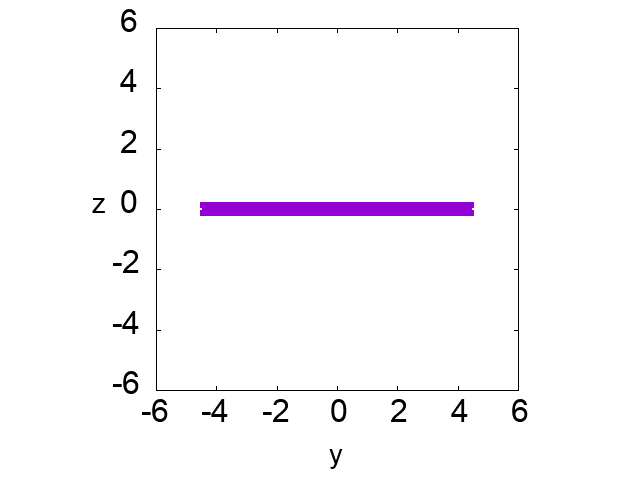}
\includegraphics[angle=0,width=5.0cm]{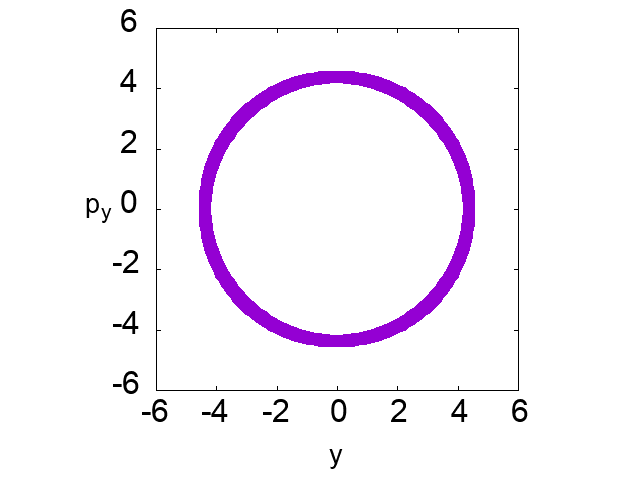}
\includegraphics[angle=0,width=5.0cm]{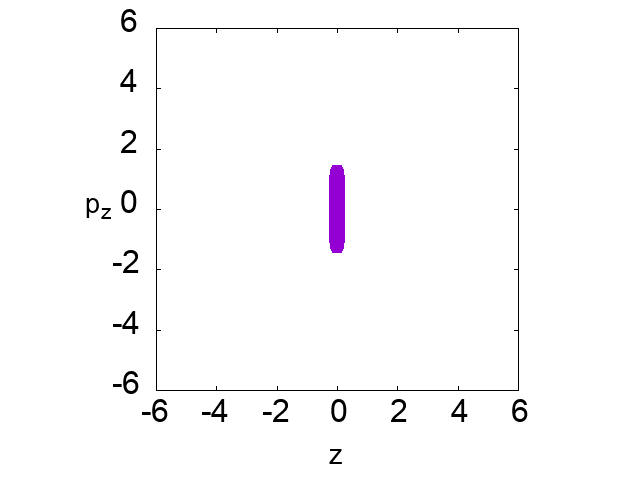}\\
\includegraphics[angle=0,width=5.0cm]{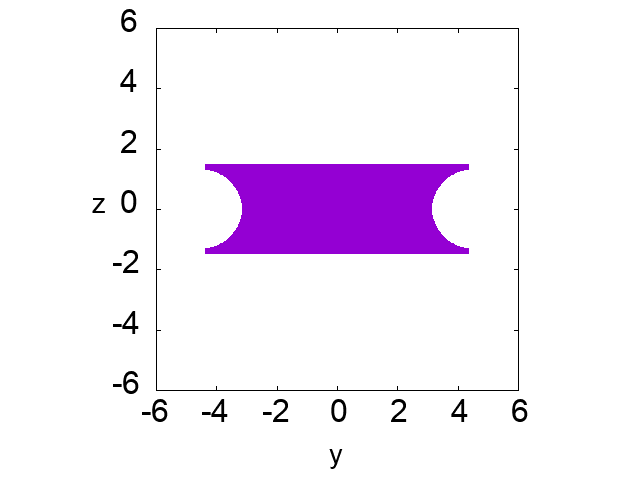}
\includegraphics[angle=0,width=5.0cm]{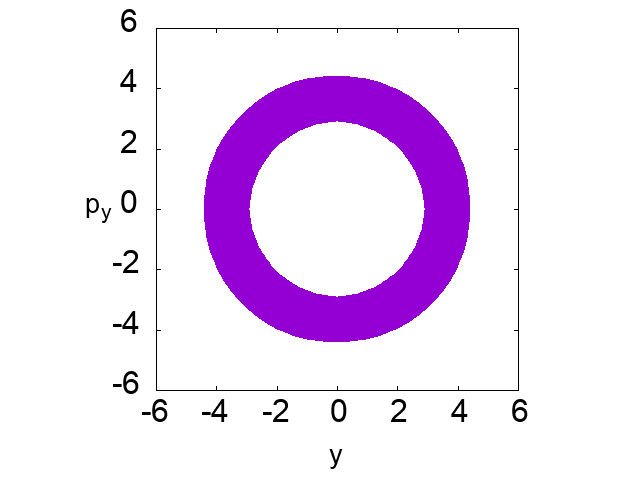}
\includegraphics[angle=0,width=5.0cm]{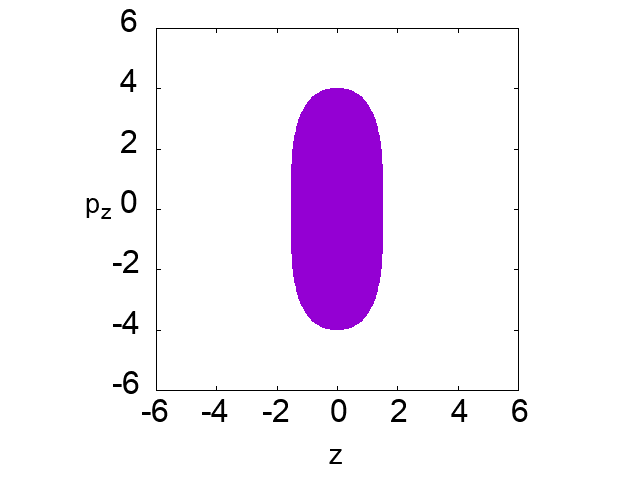}\\
\includegraphics[angle=0,width=5.0cm]{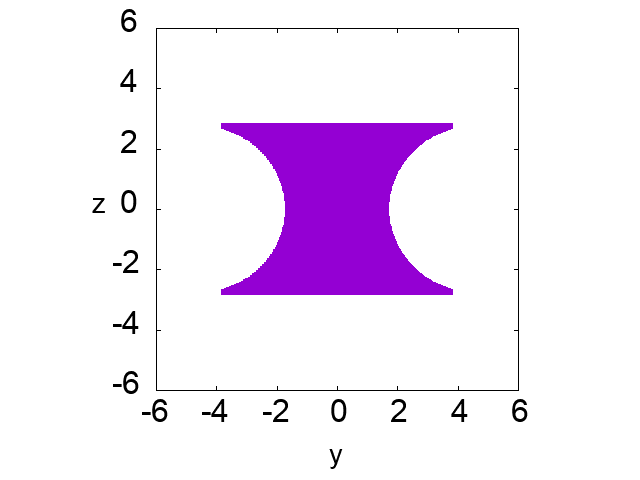}
\includegraphics[angle=0,width=5.0cm]{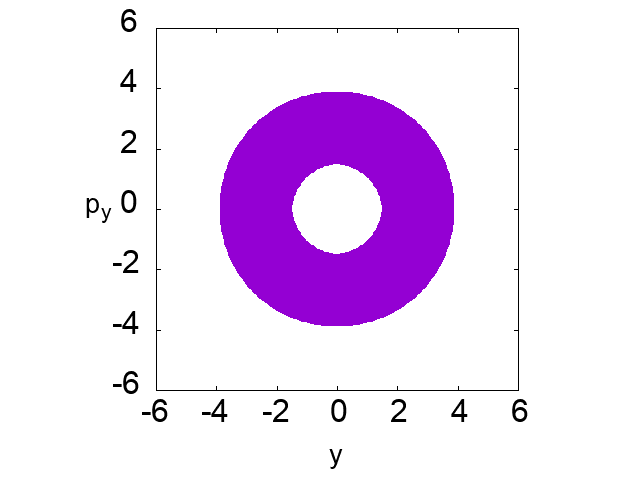}
\includegraphics[angle=0,width=5.0cm]{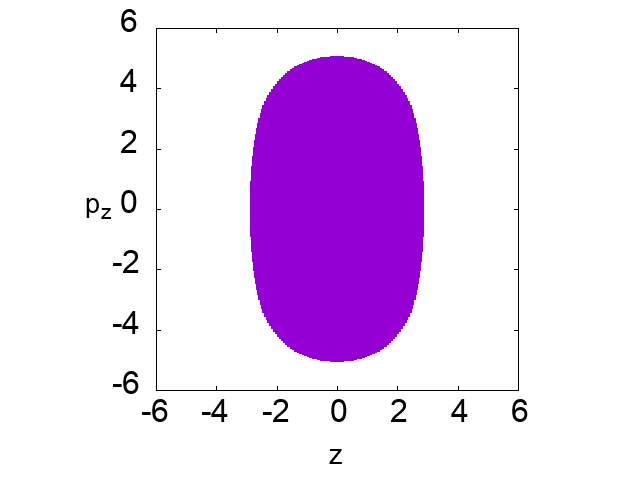}\\
\includegraphics[angle=0,width=5.0cm]{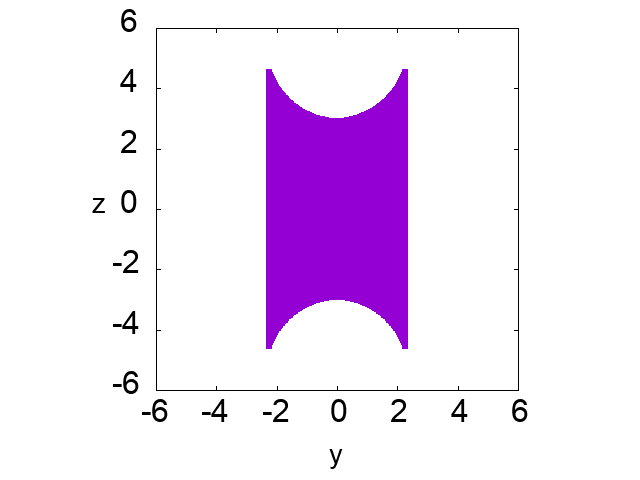}
\includegraphics[angle=0,width=5.0cm]{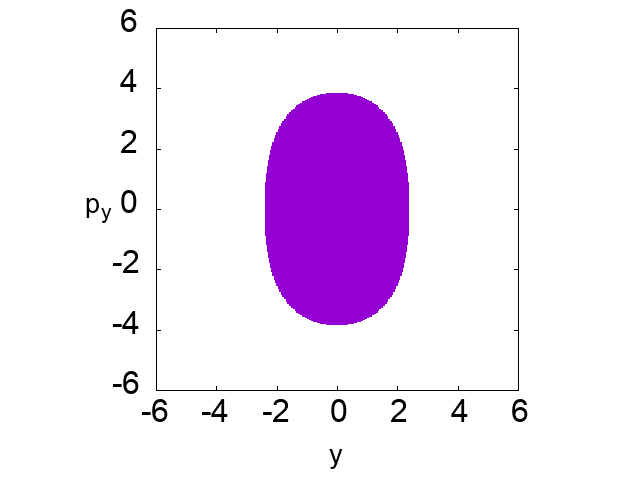}
\includegraphics[angle=0,width=5.0cm]{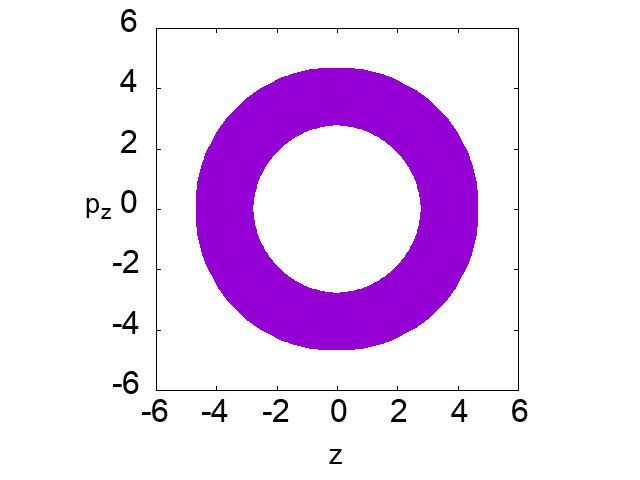}\\
\caption{2D projections of periodic orbit dividing surfaces in $(y,z)$ (first column), 
$(y,p_y)$ (second column) and $(z,p_z)$ (third column) subspaces of the phase space. The dividing surfaces  are constructed from the periodic orbit PO1 with the  associated torus  to have  r=Rmax/20 (first row),  r=Rmax/2 (second row) and r=Rmax (third row) and from the periodic PO2 with the associated torus to have r=Rmax (fourth  row).}
\label{div1}
\end{figure}

\begin{figure}
\centering
\includegraphics[angle=0,width=5.0cm]{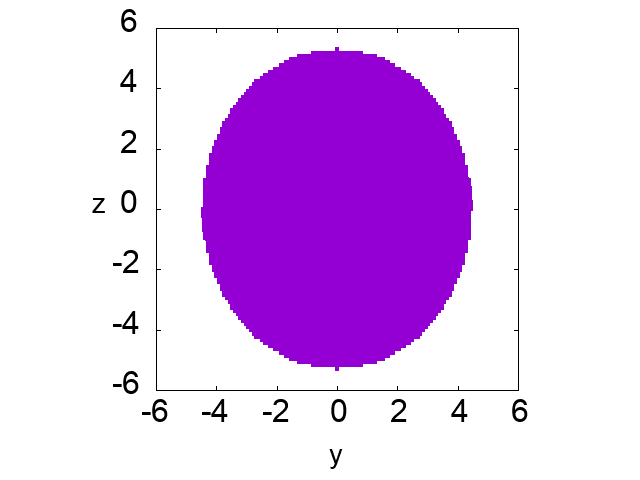}
\includegraphics[angle=0,width=5.0cm]{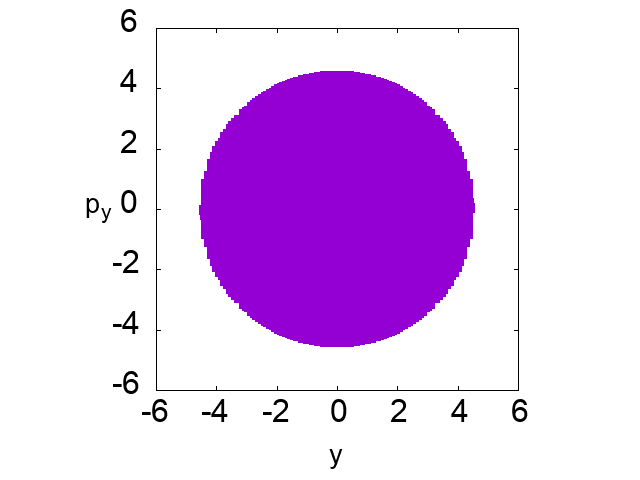}
\includegraphics[angle=0,width=5.0cm]{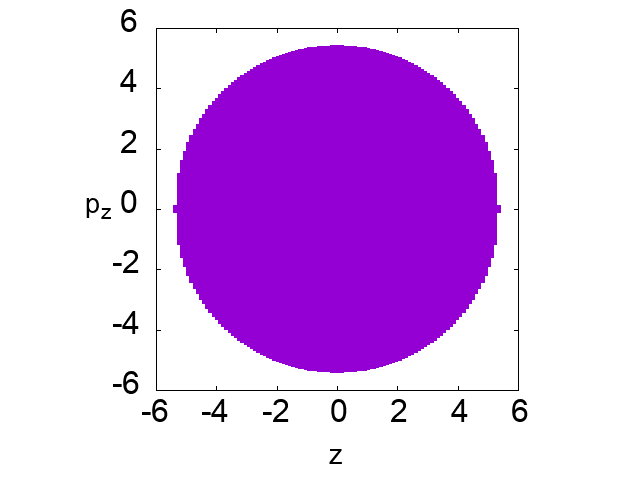}\\
\caption{2D projections of dividing surface that  is constructed from the NHIM  in $(y,z)$ (first panel), $(y,p_y)$ (second panel) and $(z,p_z)$ (third panel) subspaces of the phase space.}
\label{div1a}
\end{figure}

\begin{figure}
 \centering
\includegraphics[angle=0,width=12.0cm]{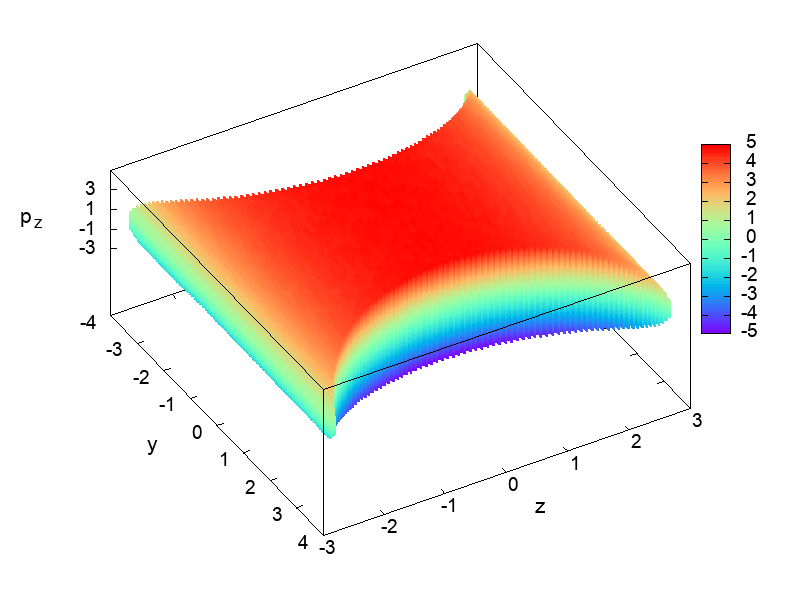}\\
\caption{ The 3D projection $(y,z,p_z)$ of the dividing surface which  is constructed from the periodic orbit PO1 with the  associated torus  to have  r=Rmax. The color indicates the values of the third dimension. The same structure is encountered  also in 3D projection  $(z,p_y,p_z)$. The viewpoint is in spherical coordinates is $(30^{o},60^{o})$. }
\label{div1-3d}
\end{figure}

\begin{figure}
 \centering
\includegraphics[angle=0,width=12.0cm]{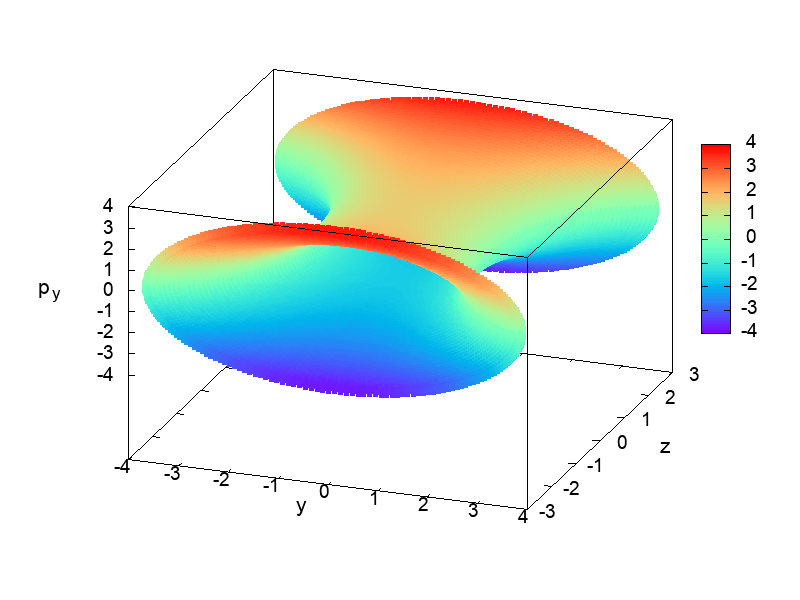}\\
\caption{ The 3D projection $(y,z,p_y)$ of the dividing surface which  is constructed from the periodic orbit PO1 with the  associated torus  to have  r=Rmax. The color indicates the values of the third dimension.The viewpoint is in spherical coordinates is $(60^{o},20^{o})$. }
\label{div2-3d}
\end{figure}

\begin{figure}
 \centering
\includegraphics[angle=0,width=12.0cm]{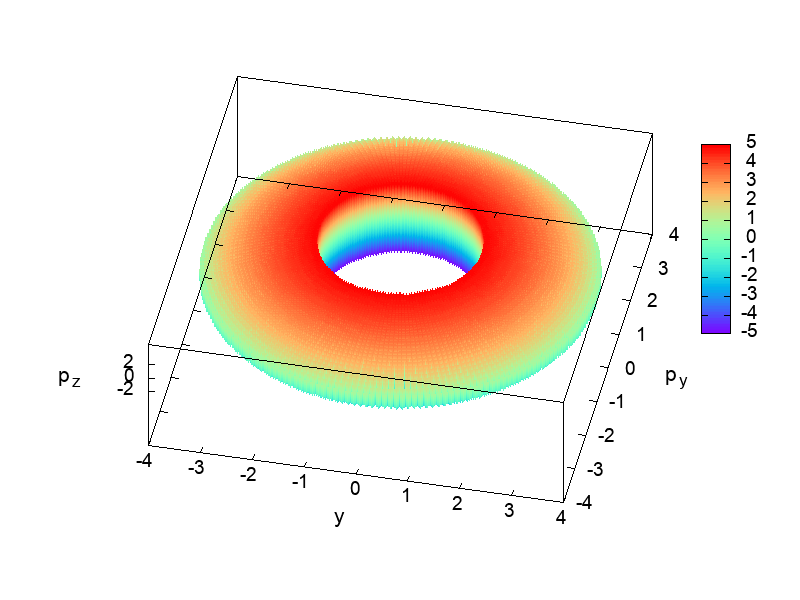}\\
\caption{ The 3D projection $(y,p_y,p_z)$ of the dividing surface which  is constructed from the periodic orbit PO1 with the  associated torus  to have  r=Rmax. The color indicates the values of the third dimension.The viewpoint is in spherical coordinates is $(20^{o},12^{o})$. }
\label{div3-3d}
\end{figure}

\begin{figure}
 \centering
\includegraphics[angle=0,width=12.0cm]{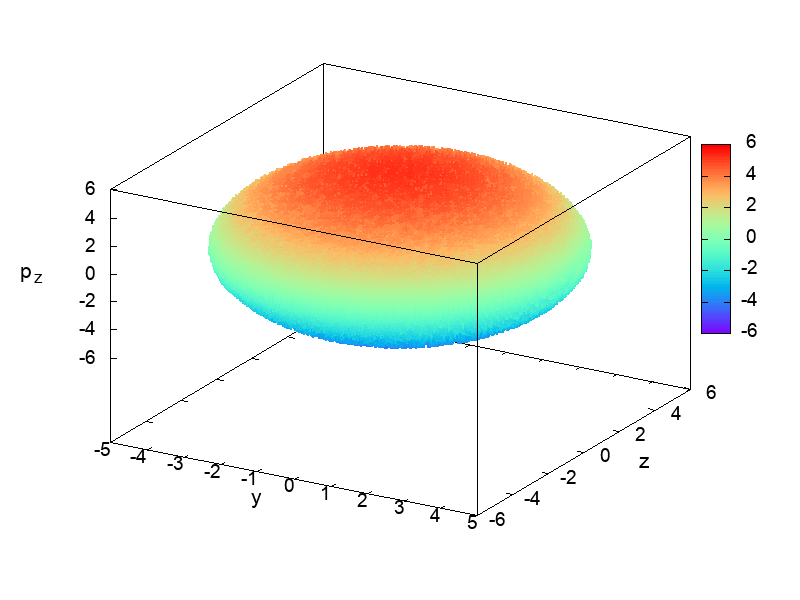}\\
\caption{ The 3D projection $(y,z,p_z)$ of the dividing surface which  is constructed from the NHIM. The color indicates the values of the third dimension.The viewpoint is in spherical coordinates is $(30^{o},60^{o})$. }
\label{div3a-3d}
\end{figure}

We studied  first the structure of these dividing surfaces and then we compare them:

\begin{enumerate}

\item \textbf{The structure of the dividing surfaces:}
All dividing surfaces that are  constructed from the periodic orbits have similar structure in the  4-dimensional space $(y,z,p_y,p_z)$. In this subsection, we choose one representative  example  of these surfaces, the dividing surface from PO1 with the  associated torus  to have  r=Rmax. We depict  this surface in all 3D projections of the 4-dimensional space $(y,z,p_y,p_z)$.  This surface is constructed as  a torus. This topology is very obvious  (as we can see in Fig. \ref{div3-3d})  in $(y,p_y,p_z)$  projection. This torus has also an hyperbolic structure that  is represented  as an  hyperboloid (as we can see in Fig. \ref{div2-3d}) in $(y,z,p_y)$ projection and hyperbolic box structure (as we can see in Fig. \ref{div1-3d}) in  $(y,z,p_z)$ and  $(z,p_y,p_z)$   projections. This means that the dividing surfaces that are constructed from periodic orbits are  hyperbolic tori 
in the  4-dimensional space $(y,z,p_y,p_z)$. 

\par The dividing surface that is constructed from the NHIM is an ellipsoid in all 3D projections of the 4-dimensional space $(y,z,p_y,p_z)$. (as we can see for example in $(y,z,p_z)$ projection in Fig.\ref{div3a-3d}). This means that this dividing surface is an ellipsoid in the 4-dimensional space.

\item \textbf{Comparison of Dividing surfaces:}
In  Fig. \ref{div1} we compare the dividing surfaces that are constructed from  the periodic orbit PO1. For this reason we use the 2D projections  $(y,z)$, $(y,p_y)$ and $(z,p_z)$. Using these projections we can understand if one dividing surface is larger in the $y$ or $z$ or $p_y$ or $p_z$ direction. This means that these  2D projections  are enough  for the comparison of different  dividing surfaces without the use of the other 2D projections 
$(y,p_z)$, $(z,p_y)$ and $(p_y,p_z)$. As we can see in the Fig. \ref{div1} the  dividing surface with associated torus with $r=Rmax/20$ has similar length in the y-direction  with the other two dividing surfaces. The dividing surfaces  form a hyperbolic box in the $(y,z)$ projection that  grows up with the increase  of the radius of the one circle  that is needed for the construction of dividing surfaces  (Fig. \ref{div1}). The increase of the radius increases  the extension of the dividing surfaces  in the z-direction. This is the reason that  the hyperbolic box of the dividing surface with  $r=Rmax/20$  is smaller than this of the    dividing surface  with $r=Rmax/2$ and this is smaller than the hyperbolic box of  the dividing surface   with   $r=Rmax$. In Fig. \ref{div1} we see that the dividing surfaces form a ring  structure in the $(y,p_y)$ projection  and an elliptic structure in the $(z,p_z)$ with increasing thickness as we increase the radius of the circle that is needed for the construction of  the dividing surfaces (see Fig. \ref{div1}). 
As we increase the radius of the circle  that generates the torus that is the basis for the construction of a periodic orbit dividing surface, this surface becomes more elongated  in  the  $p_z$  and $z$ directions (Fig. \ref{div1}).

We compare the PO2 dividing surface  with the PO1 dividing surface, both of them  associated with the torus with the maximum $r$ and $r_1$ respectively.  We compare the two cases of the dividing surfaces  in  Figs. \ref{div1}. We observe that the PO1 dividing surface is more elongated on the y-axis and the PO2 dividing surface is more elongated on the z-axis (Fig. \ref{div1}). Furthermore, the PO2 dividing surface form a ring structure in the $(z,p_z)$ and not in the $(y,p_y)$ as the PO1 dividing surface. In addition,  the PO2 dividing surface form an elliptic structure in the $(y,p_y)$ and not in the $(z,p_z)$ as the PO1 dividing surface. All these differences are because of  the fact that the PO1 is  a circle in the $(y,p_y)$ plane and PO2 is a circle in the $(z,p_z)$ plane. This means that the combination of these dividing surfaces give us the dynamical information for every direction  in the configuration space for a small area of interest (the neighbourhood of the periodic orbits).  

We compare  the dividing surface  from the NHIM with the dividing surfaces from the PO1 and PO2, both of them associated with the torus with the maximum $r$ and $r_1$ respectively. We observe in Fig. \ref{div1}  that the dividing surface  from PO1 and PO2 have the same range of values   with the dividing surface  from the NHIM (Fig.\ref{div1a})  in the $p_y$-direction and $p_z$-direction. But the PO1 dividing surface and the PO2 dividing surface have smaller range of values than this  of the dividing surface from the NHIM (Fig. \ref{div1a})  in  the $z$-directon and  $y$-direction respectively. This means  that the dividing surfaces from PO1 and PO2 are subsets of the dividing surface from the NHIM.

\end{enumerate}

\begin{figure}
 \centering
\includegraphics[angle=0,width=12.0cm]{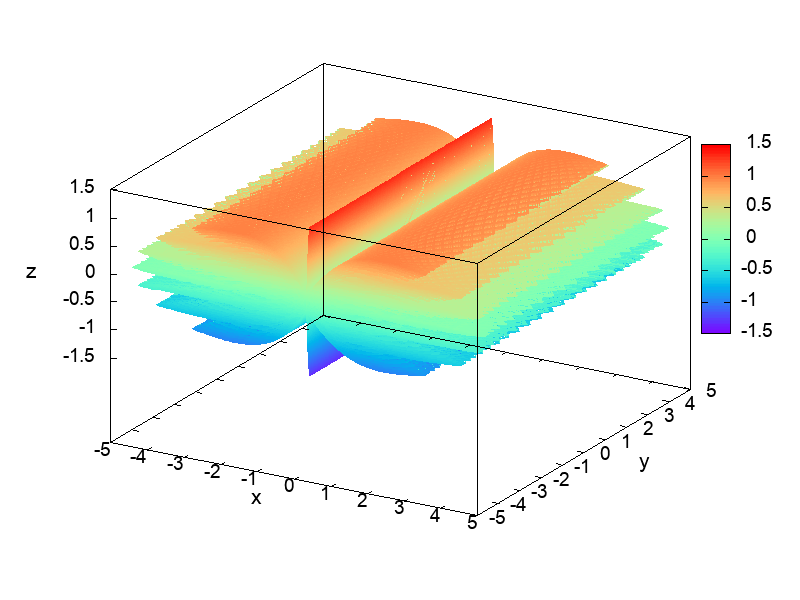}\\
\caption{ Trajectories in the 3D configuration space that   have initial conditions on the dividing surfaces which  is constructed from the periodic orbit PO1 with the  associated torus  to have  r=Rmax/20. The color indicates the values of the third dimension. The viewpoint is in spherical coordinates is $(60^{o},30^{o})$. }
\label{div2}
\end{figure}

We give an example of trajectories  that have initial conditions on one from dividing surfaces that we constructed from the periodic orbits PO1 and PO2 and we integrate them for 2 time units. We used as example  the dividing surface that is constructed from the PO1 that is constructed by a torus with small radius  $r=Rmax/20$ (Fig. \ref{div2}). We observe that this structure has as a central region  a box with a tiny  thickness and at the edges it has many sheets in the configuration space.  The structure of this surface is similar with  these that we obtained if the trajectories have initial conditions  on other dividing surfaces that are constructed from  the same or the other periodic orbit PO2.

\clearpage

\section{The Algorithm for Hamiltonian systems with $n$ degrees of freedom}
\label{sec.4}
In the  section \ref{sec.1}  we described  the difference between  the two versions of our algorithm for the construction of dividing surfaces and when we use the one or other version. Now, we describe in detail the two versions  of our algorithm in the case of Hamiltonian systems with $n$ degrees of freedom with a potential energy 
$V(x_1,x_2,...,x_n)$:

\begin{eqnarray}
\label{hgen1}
T +V(x_1,x_2,...,x_n)=E
\nonumber\\
\end{eqnarray}

\noindent
where $T$ is the kinetic energy. 

\begin{eqnarray}
\label{hgen1a}
T=p_{x_1}^2/2m_1+p_{x_2}^2/2m_2+...+p_{x_n}^2/2m_n
\nonumber\\
\end{eqnarray}

\noindent
where $p_{x_1},p_{x_2}...p_{x_n}$ are the momenta and $m_1,m_2,...m_n$ the corresponding masses.\\
When we apply the two versions  of the algorithm  to Hamiltonian systems with $n$ degrees of freedom we produce two algorithms, one algorithm for each version. 

\subsection{First Version}

The first version  of the algorithm is:

\begin{enumerate}

\item Locate an unstable periodic orbit PO for a fixed value of Energy $E$.

\item Project the PO into the configuration space and we consider  a 2D subspace of the phase space in which the projection of the periodic orbit is a closed curve (for example in the $(x_2,p_{x_2})$ space). 

\item  From the projection of the periodic orbit in the configuration space, we construct a torus that is generated by the Cartesian product  of $n-1$  circles with small radius and the projection of the periodic orbit in a 2D subspace of the phase space (for example in the $(x_2,p_{x_2})$ space). Actually topologically  it is equivalent with  the Cartesian product of $n$ circles $S^{1}\times S^{1}\times S^{1}\times...\times S^{1}$. This is a $n$-dimensional torus. This can be achieved through the  construction  of one circle around every point of the   periodic orbit in a ($n+1$)D subspace of the phase space. For example  we  compute a circle (with a fixed radius r) in the plane $(x_1,x_2)$ around every point of the periodic orbit in the  ($n+1$)D subspace  $(x_1,x_2,...,x_n,p_{x_2})$. Then we construct a new circle  around every point of the previous structure  in other 2D subspace of the  ($n+1$)D subspace $(x_1,...,x_n,p_{x_2})$. This can be done computing a circle (with a  fixed radius r) in the plane $(x_2,x_3)$. Then we continue  adding circles until we will have added $n-1$ circles to the initial projection of the periodic orbit. The target of this step is to include all coordinates of the configuration space in this torus.

\begin{eqnarray}
\label{eqt2n}
x_{2,1,i,j1}=x_{2,0,i} + r cos(\theta_{j1})
\nonumber\\
x_{1,1,i,j1}=x_{1,0,i} + r sin(\theta_{j1})
\nonumber\\
x_{3,1,i,j1}=x_{3,0,i}
\nonumber\\
...
\nonumber\\
x_{n,1,i,j1}=x_{n,0,i}
\nonumber\\
p_{x_2,1,i,j1}=p_{x_2,0,i}
\end{eqnarray}

\begin{eqnarray}
\label{eqt2na}
x_{2,2,i,j1,j2}=x_{2,1,i,j1}+r cos(\theta_{j2})
\nonumber\\
x_{3,2,i,j1,j2}=x_{3,1,i,j1}+r sin(\theta_{j2})
\nonumber\\
x_{1,2,i,j1,j2}=x_{1,1,i,j1}
\nonumber\\
...
\nonumber\\
x_{n,2,i,j1,j2}=x_{n,1,i,j1}
\nonumber\\
p_{x_2,2,i,j1,j2}=p_{x_2,1,i,j1}
\nonumber\\
...
\end{eqnarray}

\begin{eqnarray}
\label{eqt2nb}
...
\nonumber\\
x_{2,n-1,i,j1,j2,...,j(n-1)}=x_{2,n-2,i,j1,j2,...,j(n-2)}+r cos(\theta_{j(n-1)})
\nonumber\\
x_{n,n-1,i,j1,j2...j(n-1)}=x_{n,n-2,i,j1,j2,...,j(n-2)}+r sin(\theta_{j(n-1)})
\nonumber\\
x_{1,n-1,i,j1,j2,...j(n-1)}=x_{1,n-2,i,j1,j2,...,j(n-2)}
\nonumber\\
...
\nonumber\\
x_{n-1,n-1,i,j1,j2,...,j(n-1)}=x_{n-1,n-2,1,i,j1,j2,...,j(n-2)}
\nonumber\\
p_{x_2,n-1,i,j1,j2,...,j(n-1)}=p_{x_2,n-2,i,j1,j2,...,j(n-2)}
\nonumber\\
\end{eqnarray}

 $(x_{1,0,i},x_{2,0,i},...,x_{n,0,i},p_{x_2,0,i}), i=1,...N$ are  the points of the periodic orbit in the $(n+1)$D subspace $(x_1,x_2,...,x_n,p_{x_2})$. We have the angle $\theta_{j1}=j1\frac{2\pi}{n_1}$ with $j1=1,...,n_1$ for the first circle and  $\theta_{j2}=j2\frac{2k\pi}{n_1}$ with $j2=1,...,n_1$ for the second circle and so on the $\theta_{j(n-1)}=j(n-1)\frac{2\pi}{n_1}$  with $j(n-1)=1,...,n_1$  for the $n-1$ circle  that we need for the construction of the torus.

$x_{1,1,i,j1},x_{1,1,i,j1},...,x_{n,1,i,j1},p_{x_2,1,i,j1}$ with $i=1,...,N$ and $j1=1,...,n_1$  are the points of the torus that is constructed from the Cartesian product of  projection of the periodic orbit in the 2D subspace $(x_2,p_{x_2})$  and a circle in the $(x_1,x_2)$ space in the  $(n+1)$D space  $(x_1,x_2,...x_n,p_{x_2})$. $x_{2,2,i,j1,j2},x_{1,2,i,j1,j2},...x_{n,2,i,j1,j2},p_{x_2,2,i,j1,j2}$ with  $i=1,...,N$, $j1=1,...,n_1$  and $j2=1,...,n_1$ are the points of the torus that is constructed from the Cartesian product of the projection of the periodic orbit in the 2D subspace $(x_2,p_{x_2})$,  and other 2  circles  in the  ($n+1$)D space $(x_1,x_2,...,x_n,p_{x_2})$. And so on $x_{1,n-1,i,j1,j2,...j(n-1)},x_{2,n-1,i,j1,j2,...j(n-1)},...x_{n,n-1,i,j1,
j2,..j(n-1)},p_{x_2,n-1,i,j1,j2,...j(n-1)}$ with  $i=1,...,N$ and  $j1,j2,....j(n-1)=1,...,n_1$  are the points of the torus that is constructed from the Cartesian product of the projection of the periodic orbit in the 2D subspace $(x_2,p_{x_2})$,  and other $n-1$  circles  in the  ($n+1$)D space $(x_1,x_2,...,x_n,p_{x_2})$.

\item For each point $x_{1,n-1,i,j1,j2,...j(n-1)},x_{2,n-1,i,j1,j2,...j(n-1)},...x_{n,n-1,i,j1,
j2,..j(n-1)},p_{x_2,n-1,i,j1,j2,...j(n-1)}$ with  $i=1,...,N$ and  $j1,j2,....j(n-1)=1,...,n_1$ on this torus we must  calculate the $p_{x_1,n-1,i,j1,j2,,...j(n-1)}^{max}$ and $p_{x_1,n-1,i,j1,j2,,...j(n-1)}^{min}$  by  solving the following equation for  a fixed value of energy (Hamiltonian)  E with $p_{x_3}=...=p_{x_n}=0$:

\begin{eqnarray}
\label{eq6n}
V(x_{1,n-1,i,j1,j2,...j(n-1)},x_{2,n-1,i,j1,j2,...j(n-1)},...x_{n,n-1,i,j1,
j2,..j(n-1)})+
\nonumber\\
\frac{p_{x_1,n-1,i,j1,j2,...j(n-1)}^2}{2m_1}+\frac{p_{x_2,n-1,i,j1,j2,...j(n-1)}^2}{2m_2}=E
\nonumber\\
\end{eqnarray}

\noindent
and we find the maximum and minimum values \\  
$p_{x_1,n-1,i,j1,j2,...j(n-1)}^{max}$ and   $p_{x_1,n-1,i,j1,j2,...j(n-1)}^{min}$. We choose  points  $p_{x_1,n-1,i,j1,j2,...j(n-1)}$ with $j1,j2...j(n-1)=1,...,n_1$ in the interval  $p_{x_1,n-1,i,j1,j2,...j(n-1)}^{min}\leq  p_{x_1,n-1,i,j1,j2,...j(n-1)}\leq p_{x_1,n-1,i,j1,j2,...j(n-1)}^{max}$. These points can be uniformly distributed in this interval. We will repeat the same procedure  to compute the values $p_{x_3,n-1,i,j1,j2,...j(n-1)}, ...,p_{x_{n-1},n-1,i,j1,j2,...j(n-1)}$. In general for $n_2\leq n-1$ for each point $x_{1,n-1,i,j1,j2,...j(n-1)},x_{2,n-1,i,j1,j2,...j(n-1)},...x_{n,n-1,i,j1,
j2,..j(n-1)},p_{x_1,n-1,i,j1,j2,...j(n-1)},...\\
p_{x_{n_2-1},n-1,i,j1,j2,...j(n-1)}$ with  $i=1,...,N$ and  $j1,j2,....j(n-1)=1,...,n_1$  we must  calculate the $p_{x_{n_2},n-1,i,j1,j2,,...j(n-1)}^{max}$ and $p_{x_{n_2},n-1,i,j1,j2,,...j(n-1)}^{min}$  by  solving the following equation for  a fixed value of energy (Hamiltonian)  E with $p_{x_{n_2+1}}=...=p_{x_{n}}=0$:

\begin{eqnarray}
\label{eq6n1}
V(x_{1,n-1,i,j1,j2,...j(n-1)},x_{2,n-1,i,j1,j2,...j(n-1)},...x_{n,n-1,i,j1,
j2,..j(n-1)})+
\nonumber\\
\frac{p_{x_1,n-1,i,j1,j2,...j(n-1)}^2}{2m_1}+...+\frac{p_{x_{n_2},n-1,i,j1,j2,...j(n-1)}^2}{2m_{n_2}} =E
\nonumber\\
\end{eqnarray}

\noindent
and we find the maximum and minimum values   
$p_{x_{n_2},n-1,i,j1,j2,...j(n-1)}^{max}$ and   $p_{x_{n_2},n-1,i,j1,j2,...j(n-1)}^{min}$. We choose  points  $p_{x_{n_2},n-1,i,j1,j2,...j(n-1)}$ with $j1,j2,j(n-1)=1,...,n_1$ in the interval  $p_{x_{n_2},n-1,i,j1,j2,...j(n-1)}^{min}\leq  p_{x_{n_2},n-1,i,j1,j2,...j(n-1)}\leq p_{x_{n_2},n-1,i,j1,j2,...j(n-1)}^{max}$. These points can be uniformly distributed in this interval. 

\noindent
Then we obtain the value $p_{x_{n},n-1,i,j1,j2,...j(n-1)}$
from the Hamiltonian:

\begin{eqnarray}
\label{eq6n1}
V(x_{1,n-1,i,j1,j2,...j(n-1)},x_{2,n-1,i,j1,j2,...j(n-1)},...x_{n,n-1,i,j1,
j2,..j(n-1)})+
\nonumber\\
\frac{p_{x_1,n-1,i,j1,j2,...j(n-1)}^2}{2m_1}+...+\frac{p_{x_{n},n-1,i,j1,j2,...j(n-1)}^2}{2m_{n}} =E
\nonumber\\
\end{eqnarray}

\end{enumerate}

\noindent
\textit{Dimensionality and Topology}: This algorithm constructs a torus as the product of the closed curve that represents the projection of the periodic orbit (1D object) in a 2D subspace with $n-1$ circles in the ($n+1$)D subspace of the  energy manifold.  This torus is a n-dimensional torus. Then we sample the  $n-2$ variables ($n-2$ momenta)  in the interval between their  maximum and minimum value. Actually we create $n-2$ additional  segments and we increase  the dimensionality of the initial torus, from n to $2n-2$ dimensions, which is  embedded in the $2n-1$ energy surface. Then we obtain the value of the last momentum from the Hamiltonian of the system. 

\subsection{Second Version}

The second version is:

\begin{enumerate}
\item Locate an unstable periodic orbit PO for a fixed value of Energy $E$.

\item Project the PO into the configuration space and we consider  a 2D subspace of the configuration space  in which the projection of the periodic orbit is a closed curve (for example in the $(x_1,x_2)$ space). 

\item  From the projection of the periodic orbit in the configuration space, we construct a torus that is generated by the Cartesian product  of $n-2$  circles with small radius and the projection of the periodic orbit in a 2D subspace of the configuration space (for example in the $(x_1,x_2)$ space). Actually topologically  it is equivalent with  the Cartesian product of $n-1$ circles $S^{1}\times S^{1}\times S^{1}\times...\times S^{1}$. This is a $n-1$-dimensional torus. This can be achieved through the  construction  of one circle around every point of the   periodic orbit in a $n$D subspace of the phase space (configuration space). For example  we  compute a circle (with a fixed radius r) in the plane $(x_2,x_3)$ around every point of the periodic orbit in the  $n$D configuration space $(x_1,x_2,...,x_n)$. Then we construct a new circle  around every point of the previous structure in other 2D subspace of the $n$D configuration space  $(x_1,...,x_n)$. This can be done computing a circle (with a  fixed radius r) in the plane $(x_2,x_4)$. Then we continue  adding circles until we will have added $n-2$ circles to the initial projection of the periodic orbit. The target of this step is to include all coordinates of the configuration space in this torus.

\begin{eqnarray}
\label{eqt2n}
x_{2,1,i,j1}=x_{2,0,i} + r cos(\theta_{j1})
\nonumber\\
x_{3,1,i,j1}=x_{3,0,i} + r sin(\theta_{j1})
\nonumber\\
x_{1,1,i,j1}=x_{1,0,i}
\nonumber\\
x_{4,1,i,j1}=x_{4,0,i}
\nonumber\\
...
\nonumber\\
x_{n,1,i,j1}=x_{n,0,i}
\nonumber\\
\end{eqnarray}

\begin{eqnarray}
\label{eqt2na}
x_{2,2,i,j1,j2}=x_{2,1,i,j1}+r cos(\theta_{j2})
\nonumber\\
x_{4,2,i,j1,j2}=x_{4,1,i,j1}+r sin(\theta_{j2})
\nonumber\\
x_{1,2,i,j1,j2}=x_{1,1,i,j1}
\nonumber\\
x_{3,2,i,j1,j2}=x_{3,1,i,j1}
\nonumber\\
...
\nonumber\\
x_{n,2,i,j1,j2}=x_{n,1,i,j1}
\nonumber\\
...
\end{eqnarray}

\begin{eqnarray}
\label{eqt2nb}
...
\nonumber\\
x_{2,n-2,i,j1,j2,...,j(n-2)}=x_{2,n-3,i,j1,j2,...,j(n-3)}+r cos(\theta_{j(n-2)})
\nonumber\\
x_{n,n-2,i,j1,j2...j(n-2)}=x_{n,n-3,i,j1,j2,...,j(n-3)}+r sin(\theta_{j(n-2)})
\nonumber\\
x_{1,n-2,i,j1,j2,...j(n-2)}=x_{1,n-3,i,j1,j2,...,j(n-3)}
\nonumber\\
...
\nonumber\\
x_{n-1,n-2,i,j1,j2,...,j(n-2)}=x_{n-1,n-3,1,i,j1,j2,...,j(n-3)}
\nonumber\\
\end{eqnarray}

 $(x_{1,0,i},x_{2,0,i},...,x_{n,0,i}), i=1,...N$ are  the points of the periodic orbit in the $n$D configuration space $(x_1,x_2,...,x_n)$. We have the angle $\theta_{j1}=j1\frac{2\pi}{n_1}$ with $j1=1,...,n_1$ for the first circle and  $\theta_{j2}=j2\frac{2k\pi}{n_1}$ with $j2=1,...,n_1$ for the second circle and so on the $\theta_{j(n-2)}=j(n-2)\frac{2\pi}{n_1}$  with $j(n-2)=1,...,n_1$  for the $n-2$ circle  that we need for the construction of the torus.

$x_{1,1,i,j1},x_{2,1,i,j1},...,x_{n,1,i,j1}$ with $i=1,...,N$ and $j1=1,...,n_1$  are the points of the torus that is constructed from the Cartesian product of the projection of the periodic orbit in the 2D subspace $(x_1,x_2)$  and a circle in the  $n$D space  $(x_1,x_2,...x_n)$. $x_{1,2,i,j1,j2},x_{2,2,i,j1,j2},...x_{n,2,i,j1,j2}$ with  $i=1,...,N$, $j1=1,...,n_1$  and $j2=1,...,n_1$ are the points of the torus that is constructed from the Cartesian product of the projection of the periodic orbit in the 2D subspace $(x_1,x_2)$,  and other 2  circles in the  $n$D space $(x_1,x_2,...,x_n)$. And so on $x_{1,n-2,i,j1,j2,...j(n-2)},x_{2,n-2,i,j1,j2,...j(n-2)},...x_{n,n-2,i,j1,
j2,..j(n-2)}$ with  $i=1,...,N$ and  $j1,j2,....j(n-2)=1,...,n_1$  are the points of the torus that is constructed from the Cartesian product of the projection of the periodic orbit in the 2D subspace $(x_1,x_2)$,  and other $n-2$  circles  in the  $n$D space $(x_1,x_2,...,x_n)$.

\item For each point $x_{1,n-2,i,j1,j2,...j(n-2)},x_{2,n-2,i,j1,j2,...j(n-2)},...x_{n,n-2,i,j1,
j2,..j(n-2)}$ with  $i=1,...,N$ and  $j1,j2,....j(n-2)=1,...,n_1$ on this torus we must  calculate the $p_{x_1,n-2,i,j1,j2,,...j(n-2)}^{max}$ and $p_{x_1,n-2,i,j1,j2,,...j(n-2)}^{min}$  by  solving the following equation for  a fixed value of energy (Hamiltonian)  E with $p_{x_2}=p_{x_3}=...=p_{x_n}=0$:

\begin{eqnarray}
\label{eq6n}
V(x_{1,n-2,i,j1,j2,...j(n-2)},x_{2,n-2,i,j1,j2,...j(n-2)},...x_{n,n-2,i,j1,
j2,..j(n-2)})+
\nonumber\\
\frac{p_{x_1,n-2,i,j1,j2,...j(n-2)}^2}{2m_1}=E
\nonumber\\
\end{eqnarray}

\noindent
and we find the maximum and minimum values \\  
$p_{x_1,n-2,i,j1,j2,...j(n-2)}^{max}$ and   $p_{x_1,n-2,i,j1,j2,...j(n-2)}^{min}$. We choose  points  $p_{x_1,n-2,i,j1,j2,...j(n-2)}$ with $j1,j2...j(n-2)=1,...,n_1$ in the interval  $p_{x_1,n-2,i,j1,j2,...j(n-2)}^{min}\leq  p_{x_1,n-2,i,j1,j2,...j(n-2)}\leq p_{x_1,n-2,i,j1,j2,...j(n-2)}^{max}$. These points can be uniformly distributed in this interval. We will repeat the same procedure  to compute the values $p_{x_2,n-2,i,j1,j2,...j(n-2)}, ...,p_{x_{n-1},n-2,i,j1,j2,...j(n-2)}$. In general for $n_2\leq n-1$ for each point $x_{1,n-2,i,j1,j2,...j(n-2)},x_{2,n-2,i,j1,j2,...j(n-2)},...x_{n,n-2,i,j1,
j2,..j(n-2)},p_{x_1,n-2,i,j1,j2,...j(n-2)},...\\
p_{x_{n_2-1},n-2,i,j1,j2,...j(n-2)}$ with  $i=1,...,N$ and  $j1,j2,....j(n-2)=1,...,n_1$  we must  calculate the $p_{x_{n_2},n-2,i,j1,j2,,...j(n-2)}^{max}$ and $p_{x_{n_2},n-2,i,j1,j2,,...j(n-2)}^{min}$  by  solving the following equation for  a fixed value of energy (Hamiltonian)  $E$ with $p_{x_{n_2+1}}=...=p_{x_{n}}=0$:

\begin{eqnarray}
\label{eq6n1}
V(x_{1,n-2,i,j1,j2,...j(n-2)},x_{2,n-2,i,j1,j2,...j(n-2)},...x_{n,n-2,i,j1,
j2,..j(n-2)})+
\nonumber\\
\frac{p_{x_1,n-2,i,j1,j2,...j(n-2)}^2}{2m_1}+...+\frac{p_{x_{n_2},n-2,i,j1,j2,...j(n-2)}^2}{2m_{n_2}} =E
\nonumber\\
\end{eqnarray}

and we find the maximum and minimum values   
$p_{x_{n_2},n-2,i,j1,j2,...j(n-2)}^{max}$ and   $p_{x_{n_2},n-2,i,j1,j2,...j(n-2)}^{min}$. We choose  points  $p_{x_{n_2},n-2,i,j1,j2,...j(n-2)}$ with $j1,j2,j(n-2)=1,...,n_1$ in the interval  $p_{x_{n_2},n-2,i,j1,j2,...j(n-2)}^{min}\leq  p_{x_{n_2},n-2,i,j1,j2,...j(n-2)}\leq p_{x_{n_2},n-2,i,j1,j2,...j(n-2)}^{max}$. These points can be uniformly distributed in this interval. 

Then we obtain the value $p_{x_{n},n-2,i,j1,j2,...j(n-2)}$
from the Hamiltonian:

\begin{eqnarray}
\label{eq6n1}
V(x_{1,n-2,i,j1,j2,...j(n-2)},x_{2,n-2,i,j1,j2,...j(n-2)},...x_{n,n-2,i,j1,
j2,..j(n-2)})+
\nonumber\\
\frac{p_{x_1,n-2,i,j1,j2,...j(n-2)}^2}{2m_1}+...+\frac{p_{x_{n},n-2,i,j1,j2,...j(n-2)}^2}{2m_n} =E
\nonumber\\
\end{eqnarray}

\end{enumerate}

\noindent
\textit{Dimensionality and Topology}: This algorithm constructs a torus as the product of the closed curve that represents the projection of the periodic orbit (1D object) in a 2D subspace  of the configuration space with $n-2$ circles in the $n$D  configuration space .  This torus is a $n-1$-dimensional torus. Then we sample the  $n-1$ variables ($n-1$ momenta)  in the interval between their  maximum and minimum value. Actually we create  $n-1$ additional segments and we increase  the dimensionality of the initial torus, from $n-1$ to $2n-2$ dimensions, which is  embedded in the $2n-1$ energy surface. Then we obtain the value of the last momentum from the Hamiltonian of the system.

\section{Conclusions}
\label{concl}

We generalized the notion of the periodic orbit dividing surface construction  to Hamiltonian systems with $n$ degrees of freedom. This is very important because until now this surface was computed only in Hamiltonian systems with two degrees of freedom though the classical method of \cite{pechukas1981} and \cite{pollak1985}. For Hamiltonian systems with three or more degrees of freedom, the dividing surfaces could be computed using  as a starting point the Normally Hyperbolic Invariant Manifold (NHIM- see \cite{wiggins2016}, \cite{wiggins1994}) and not periodic orbits. In many systems this is very difficult and we need the Normal form theory  to compute this structure.  In this method we use the periodic orbit as a starting point  and  we construct  a torus  as a Cartesian  product of $n-1$ circle or $n-2$ circles and a 2D projection of the periodic orbit avoiding the difficult computation of the NHIM.  We compare the dividing surfaces  from the periodic   orbits with the dividing surface from the NHIM in a simple Normal Form quadratic Hamiltonian system with two and three degrees of freedom. From all these we have the following remarks:

\begin{enumerate} 

\item Our algorithm for the construction of the periodic orbit dividing surfaces is valid in the cases in which the periodic orbits are closed curves  in a 2D subspace of the configuration space or a 2D subspace of the extended configuration space (a space that consists of the configuration space plus 1D space that corresponds to one of the momenta).

\item We have two versions of our algorithm for the construction of the dividing surfaces from periodic orbits. The choice of one of these versions depends from the fact if the  periodic orbit is closed curve in a 2D subspace of the configuration space or not.

\item In Hamiltonian systems with two degrees of freedom  the second version  of our algorithm for the construction of dividing surfaces coincides with the classical method of \cite{pechukas1981} and \cite{pollak1985}.

\item In Hamiltonian systems with three degrees of freedom, the periodic orbit dividing surfaces have the topology of a hyperbolic torus. 

\item The periodic orbit dividing surfaces that are constructed from our algorithm are subsets of the dividing surfaces that are constructed from the NHIM.

\item According to the position of the periodic orbits in the phase space the associated dividing surfaces, that are constructed from our algorithm,  give us  the trajectory behaviour for different regions of the configuration space. For example in our case (a Normal form Hamiltonian system with three degrees of freedom) the PO1 dividing surface give us more information for the trajectories in the y-direction and PO2 dividing surface  give us more information  in the z-direction.

\item The radius of the circles that are needed for the  construction of the periodic orbit dividing surfaces, through our algorithm,  is crucial for the efficiency of the algorithm and the size of the dividing surfaces.

\end{enumerate}

\nonumsection{Acknowledgments}We acknowledge the support of EPSRC Grant No.~EP/P021123/1 and ONR Grant No.~N00014-01-1-0769.

\bibliographystyle{ws-ijbc}
\bibliography{caldera2c}

\end{document}